\documentclass[twoside,twocolumn,9pt]{article}
\usepackage{extsizes}
\usepackage[super,sort&compress,comma]{natbib} 
\usepackage[version=3]{mhchem}
\usepackage[left=1.5cm, right=1.5cm, top=1.785cm, bottom=2.0cm]{geometry}
\usepackage{balance}
\usepackage{mathptmx}
\usepackage{sectsty}
\usepackage{graphicx} 
\usepackage{lastpage}
\usepackage[format=plain,justification=justified,singlelinecheck=false,font={stretch=1.125,small,sf},labelfont=bf,labelsep=space]{caption}
\usepackage{float}
\usepackage{fancyhdr}
\usepackage{fnpos}
\usepackage[english]{babel}
\addto{\captionsenglish}{%
  \renewcommand{\refname}{Notes and references}
}
\usepackage{array}
\usepackage{droidsans}
\usepackage{charter}
\usepackage[T1]{fontenc}
\usepackage[usenames,dvipsnames]{xcolor}
\usepackage{setspace}
\usepackage[compact]{titlesec}
\usepackage{hyperref}

\usepackage{epstopdf}

\usepackage{subcaption}
\usepackage{makecell}

\definecolor{cream}{RGB}{222,217,201}

\begin{document}

\pagestyle{fancy}
\thispagestyle{plain}
\fancypagestyle{plain}{
\renewcommand{\headrulewidth}{0pt}
}

\makeFNbottom
\makeatletter
\renewcommand\LARGE{\@setfontsize\LARGE{15pt}{17}}
\renewcommand\Large{\@setfontsize\Large{12pt}{14}}
\renewcommand\large{\@setfontsize\large{10pt}{12}}
\renewcommand\footnotesize{\@setfontsize\footnotesize{7pt}{10}}
\makeatother

\renewcommand{\thefootnote}{\fnsymbol{footnote}}
\renewcommand\footnoterule{\vspace*{1pt}%
\color{cream}\hrule width 3.5in height 0.4pt \color{black}\vspace*{5pt}} 
\setcounter{secnumdepth}{5}

\makeatletter 
\renewcommand\@biblabel[1]{#1}            
\renewcommand\@makefntext[1]%
{\noindent\makebox[0pt][r]{\@thefnmark\,}#1}
\makeatother 
\renewcommand{\figurename}{\small{Fig.}~}
\sectionfont{\sffamily\Large}
\subsectionfont{\normalsize}
\subsubsectionfont{\bf}
\setstretch{1.125} 
\setlength{\skip\footins}{0.8cm}
\setlength{\footnotesep}{0.25cm}
\setlength{\jot}{10pt}
\titlespacing*{\section}{0pt}{4pt}{4pt}
\titlespacing*{\subsection}{0pt}{15pt}{1pt}

\fancyfoot{}
\fancyfoot[LO,RE]{\vspace{-7.1pt}\includegraphics[height=9pt]{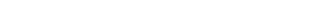}}
\fancyfoot[CO]{\vspace{-7.1pt}\hspace{11.9cm}\includegraphics{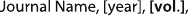}}
\fancyfoot[CE]{\vspace{-7.2pt}\hspace{-13.2cm}\includegraphics{head_foot/RF}}
\fancyfoot[RO]{\footnotesize{\sffamily{1--\pageref{LastPage} ~\textbar  \hspace{2pt}\thepage}}}
\fancyfoot[LE]{\footnotesize{\sffamily{\thepage~\textbar\hspace{4.65cm} 1--\pageref{LastPage}}}}
\fancyhead{}
\renewcommand{\headrulewidth}{0pt} 
\renewcommand{\footrulewidth}{0pt}
\setlength{\arrayrulewidth}{1pt}
\setlength{\columnsep}{6.5mm}
\setlength\bibsep{1pt}

\makeatletter 
\newlength{\figrulesep} 
\setlength{\figrulesep}{0.5\textfloatsep} 

\newcommand{\topfigrule}{\vspace*{-1pt}%
\noindent{\color{cream}\rule[-\figrulesep]{\columnwidth}{1.5pt}} }

\newcommand{\botfigrule}{\vspace*{-2pt}%
\noindent{\color{cream}\rule[\figrulesep]{\columnwidth}{1.5pt}} }

\newcommand{\dblfigrule}{\vspace*{-1pt}%
\noindent{\color{cream}\rule[-\figrulesep]{\textwidth}{1.5pt}} }

\makeatother

\twocolumn[
  \begin{@twocolumnfalse}
{\includegraphics[height=30pt]{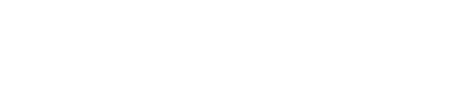}\hfill\raisebox{0pt}[0pt][0pt]{\includegraphics[height=55pt]{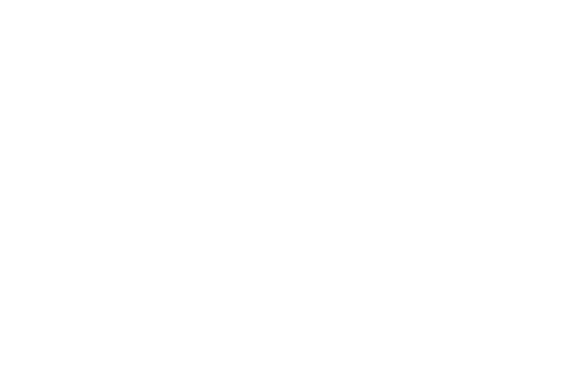}}\\[1ex]
\includegraphics[width=18.5cm]{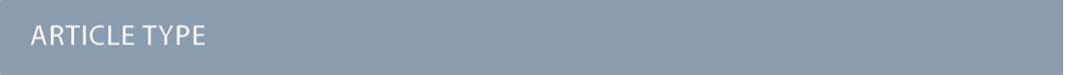}}\par
\vspace{1em}
\sffamily
\begin{tabular}{m{4.5cm} p{13.5cm} }

\includegraphics{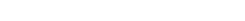} & \noindent\LARGE{\textbf{Computational study of \ce{Li+} solvation structures in fluorinated ether, non-fluorinated ether, and organic carbonate-based electrolytes at low and high salt concentrations$^\dag$}} \\
\vspace{0.3cm} & \vspace{0.3cm} \\

& \noindent\large{Rumana Hasan$^{\ast}$ and Dibakar Datta$^{\ast}$} \\

\includegraphics{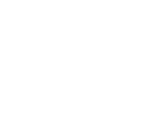} & \noindent\normalsize{Understanding the solvation structure of electrolytes is crucial for optimizing the performance and stability of lithium-ion batteries. Novel electrolytes are essential for enhancing electrolyte structure and ensuring better integration with modern electrode systems. Herein, we report a novel weakly solvated ether electrolyte (WSEE) composed of a pure fluorinated ether solvent, which results in an anion-rich solvation structure even at a low salt concentration of 1 M. To explore this, we selected the advanced fluorinated solvent 2,2-difluoroethyl methyl ether (FEME) and compared it with dipropyl ether (DPE), ethylene carbonate (EC), and diethyl carbonate (DEC). The prepared electrolyte systems include DPE with 1 M, 1.8 M, and 4 M \ce{LiFSI}; FEME with 1 M, 1.8 M, and 4 M \ce{LiFSI}; and a 1:1 vol\% EC/DEC mixture containing 1 M \ce{LiPF6}. In this work, we comprehensively investigate the \ce{Li+} solvation structures using molecular dynamics (MD) simulations and density functional theory (DFT) calculations. Our computational findings indicate the presence of large ion aggregates (AGGs) in each DPE- and FEME-based electrolyte, while SSIPs (68\%) are the dominant species in the mixed EC/DEC electrolyte. Notably, the formation of large ion aggregates is more pronounced in FEME-based electrolytes. The dominant solvation structures in the ether-based electrolytes are the anion-rich complexes \ce{Li+\ce{(FSI-})3(DPE)1} and \ce{Li+\ce{(FSI-})3(FEME)1}. We find that, similar to DPE, the FEME solvent also exhibits weak solvating power across all examined salt concentrations. More specifically, we find that FEME has weaker solvating power than DPE. This behavior is predicted by MD simulations, which indicate a strong preference for \ce{Li+} ions to coordinate with \ce{FSI-} anions within the primary solvation shell. We also observe that the number of unique solvation structures in the ether-based electrolytes increases with salt concentration, with FEME+\ce{LiFSI} showing slightly more unique solvation structures than DPE+\ce{LiFSI}. Furthermore, the quantum mechanical features of the \ce{Li+} solvation structures in DPE+1.8 M \ce{LiFSI}, FEME+1.8 M \ce{LiFSI}, and EC/DEC+1 M \ce{LiPF6} electrolytes are analyzed in detail using DFT calculations. We anticipate that this study will provide valuable insights into the \ce{Li+} solvation structures in DPE, FEME, and EC/DEC electrolytes, where the ether-based electrolytes exhibit closely similar properties.} \\

\end{tabular}

 \end{@twocolumnfalse} \vspace{0.6cm}

  ]

\renewcommand*\rmdefault{bch}\normalfont\upshape
\rmfamily
\section*{}
\vspace{-1cm}


\footnotetext{\textit{Department of Mechanical and Industrial Engineering, New Jersey Institute of Technology, Newark, NJ 07102, USA. E-mail: rh432@njit.edu, dibakar.datta@njit.edu}}


\footnotetext{\dag~Electronic supplementary information (ESI) available. See DOI: 10.1039/cXCP00000x/}







\section{Introduction}

\begin{figure*}[htbp]
\centering
\begin{minipage}{.5\textwidth}
  \centering
  \includegraphics[width=\linewidth]{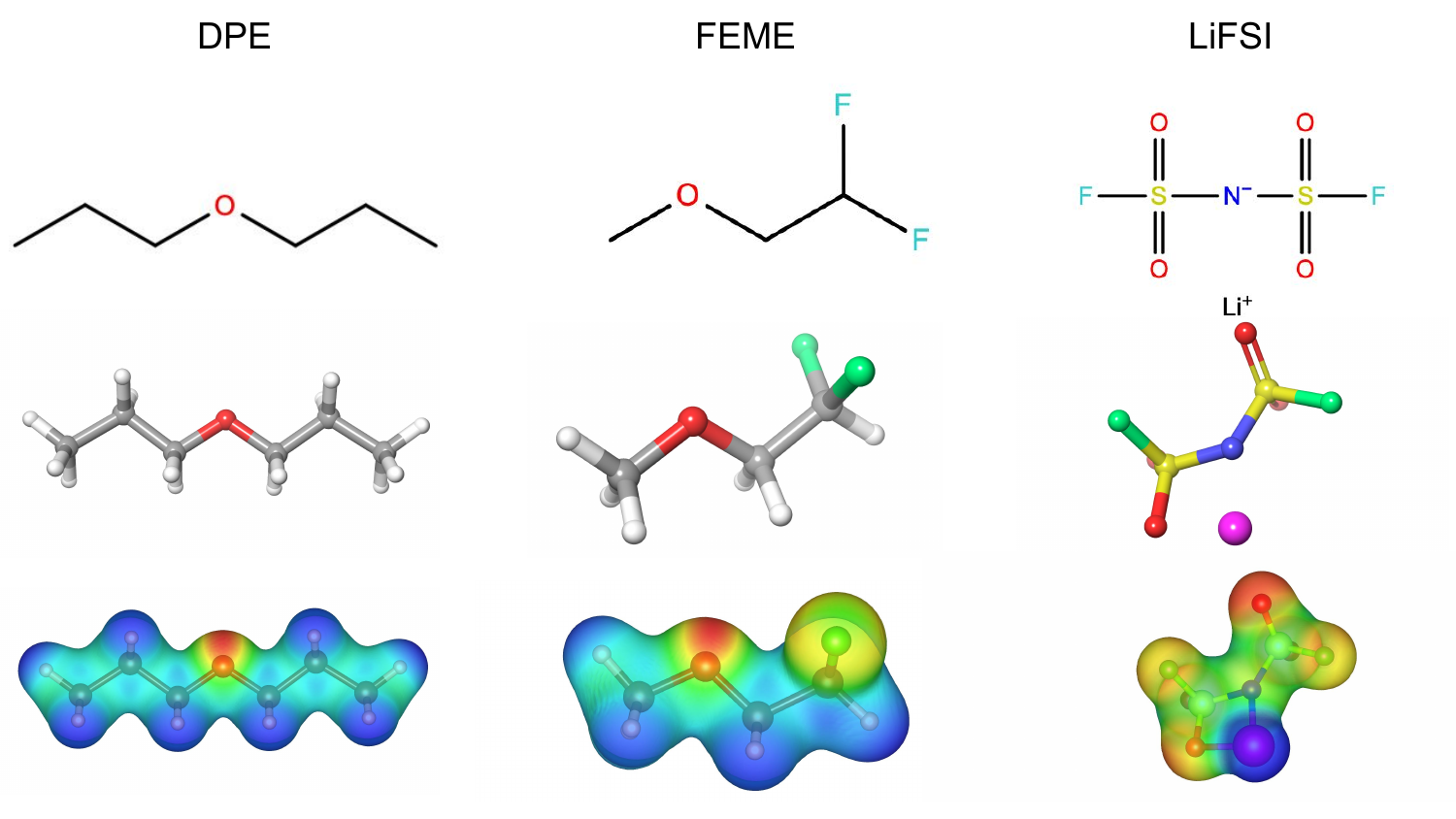}
  \label{fig:DPE_FEME_ESP}
\end{minipage}%
\hfill
\begin{minipage}{.5\textwidth}
  \centering
  \includegraphics[width=\linewidth]{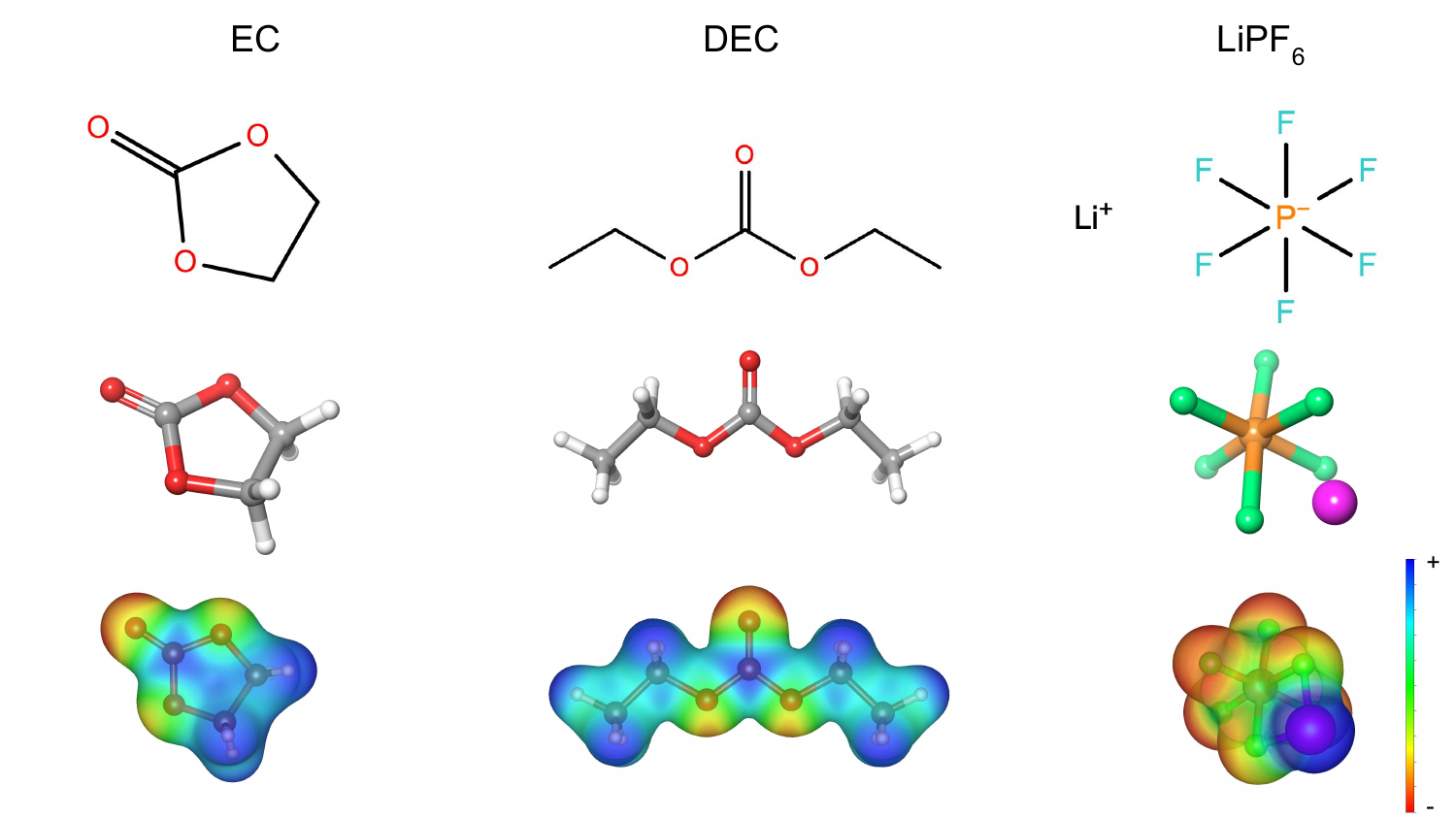}
  \label{fig:ECDEC_ESP}
\end{minipage}%
\caption{2D and 3D structures with ESP maps of the studied salts and solvents. The red and blue regions represent areas of high electron density (negative charge) and low electron density (positive charge), respectively. The magenta, green, orange, red, gray/brown, white/light pink, yellow, and blue spheres represent Li, F, P, O, C, H, S, and N atoms, respectively. For all ESP maps, the isosurface level was set at 10 eV. The isosurfaces of ESP maps were visualized using VESTA.}
\label{fig:compact_figures_ESP}
\end{figure*}

Rechargeable lithium-ion batteries (LIBs) are essential to modern energy storage, powering applications from consumer electronics to electric vehicles due to their high energy density, exceeding 300 Wh/kg \cite{xu2004nonaqueous, armand2008building, goodenough2010challenges, tarascon2001issues, scrosati2011lithium, li201830, wu2020empirical}, long cycle life, and stability. Apart from electrode materials, choosing an appropriate electrolyte to facilitate lithium-ion transport between electrodes is a complex task that demands a deep understanding of the electrolyte structure \cite{goodenough2010challenges}. As the demand for longer-lasting and safer batteries grows, optimizing the electrolyte has become a critical research focus \cite{xu2023electrolyte}. To achieve this, frameworks are currently being developed to enable efficient searches for electrolyte materials \cite{cheng2015accelerating, qu2015electrolyte, borodin2015towards}. The structure and dynamics of the electrolyte, particularly the solvation environment around \ce{Li+}, play a vital role in determining key battery properties, including ion conductivity, electrochemical stability, and solid-electrolyte interphase (SEI) layer formation on the electrodes \cite{bieker2015electrochemical, tsai2021effect, baek2021photochemically}. The transport mechanism of these \ce{Li+} ions within the electrolyte depends on their specific solvation structure, which is defined by the coordination of solvent molecules and anions around the \ce{Li+} ion. Therefore, a thorough understanding of the solvation structure is important for the development of improved electrolytes. Various analytical and computational techniques, including FTIR \cite{chae2022lithium, lim2019two, fulfer2017comparison, lee2017ultrafast, fulfer2016solvation, seo2015role, chapman2017spectroscopic}, Raman \cite{cresce2017solvation, hwang2018ionic, lee2020does}, NMR spectroscopy \cite{seo2015role, chapman2017spectroscopic, cresce2017solvation, bogle2013understanding}, DFT calculations \cite{fulfer2016solvation, seo2015role, chapman2017spectroscopic, skarmoutsos2015li+, borodin2016competitive, wang2024situ}, and MD simulations \cite{wróbel2021metfsi, borodin2009quantum, liang2017revealing, han2019structure, mynam2019molecular, zhang2019ab} have been utilized to investigate solvation structures.

\begin{figure*}[h]
 \centering
 \includegraphics[width=\textwidth]{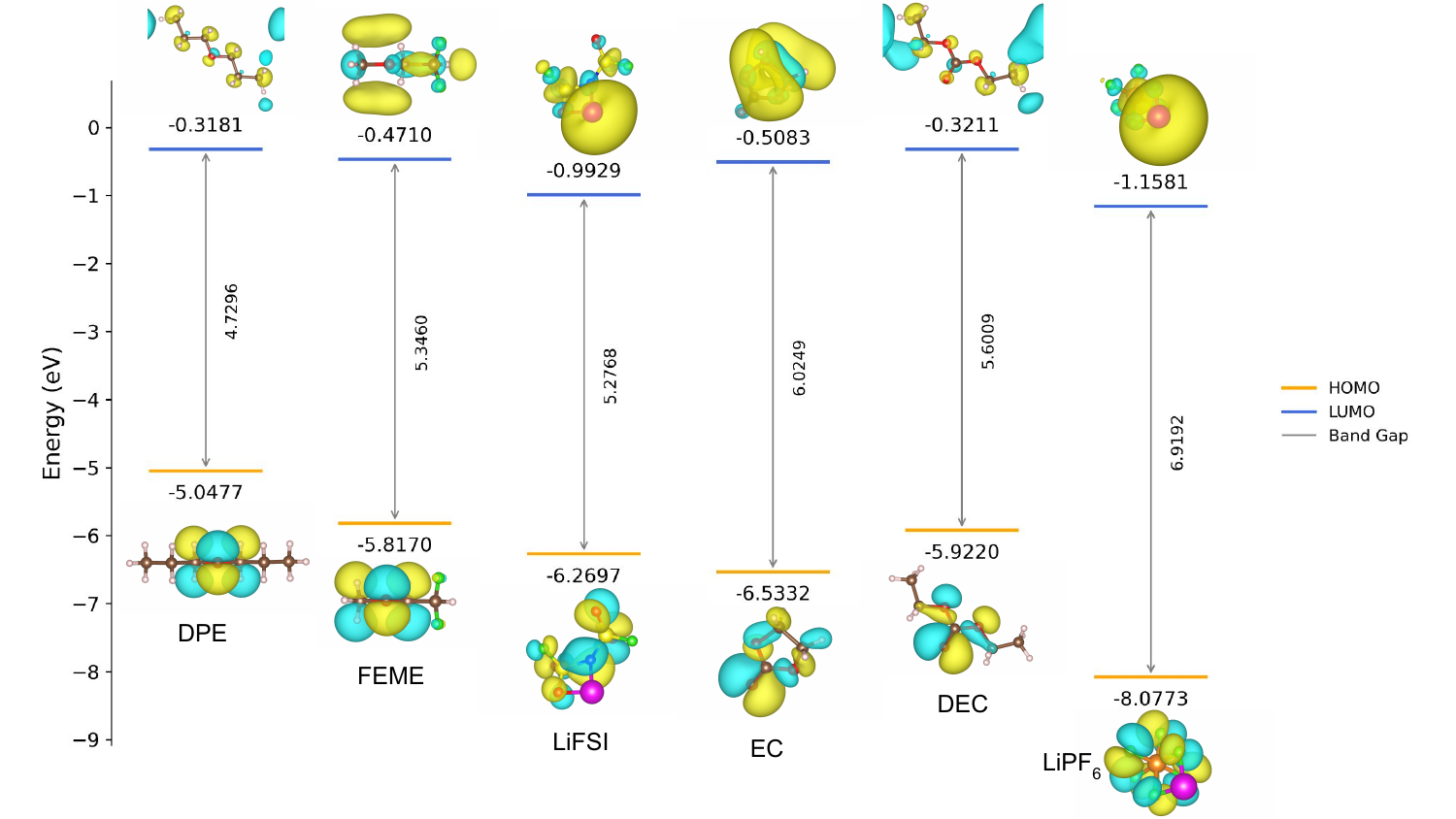}
 \caption{Comparison of HOMO/LUMO energy levels of the studied salts and solvents. The positive and negative phase of HOMO and LUMO are depicted in yellow and cyan colors, respectively. Yellow and cyan indicate the different signs of the isosurface of the wave function, and their sizes indicate its amplitude. For all HOMO/LUMO diagrams, the isosurface level was set between $8 \times 10^{-9}$ and $8 \times 10^{-8}$ $e/bohr^3$. The isosurfaces of HOMO and LUMO were visualized using VESTA.}
 \label{fig:HOMOLUMO}
\end{figure*}

As mentioned above, the movement of \ce{Li+} ions through the electrolyte is recognized as a crucial factor influencing the rate at which energy is transferred to the electrodes \cite{boyer2016structure}. According to the literature, \ce{Li+} ion transport occurs in two stages: first, the \ce{Li+} ions become surrounded by solvent molecules, and then these solvated ions migrate \cite{hou2021solvation}. Gaining deeper insights into the solvation and transport behavior of \ce{Li+} ions can enable the development of improved electrolytes. Recent studies highlight the importance of the solvation structure of electrolytes and its impact on battery performance \cite{cheng2022emerging, piao2023review}. Specifically, weakly solvated ether electrolytes (WSEEs) have been shown to exhibit anion-rich solvation structures, which have attracted significant attention \cite{yao2021regulating, pham2021simultaneous}. In contrast, carbonate-based electrolytes, widely used in commercial LIBs, often feature solvent-separated ion pairs (SSIPs) as the dominant solvation species due to their high solvation power \cite{hou2021solvation, chen2020electrolyte}. This solvating power of a solvent is determined by the strength of the ion–dipole interaction between solvent molecules and \ce{Li+} ions. Chen et al. observed that the solvating power is governed by several factors, including the dipole moment and molecular orientation of the solvent, donor number, the extent of competition between solvents and anions in coordinating with \ce{Li+}, and the dielectric constant \cite{chen2023correlating}. The study also showed that the competition between \ce{Li+-solvent} and \ce{Li+-anion} interactions largely determines the final \ce{Li+} solvation structures. However, Su et al. observed that the solvating power of solvents is primarily governed by their molecular structure, including steric hindrance and coordination ability, rather than by dielectric constant or donor number \cite{su2019solvating}. Earlier studies have further demonstrated that no single physical parameter, such as dielectric constant or dipole moment, can fully describe solvating power \cite{reichardt2021solvation}. Chen et al. also noted that there is currently no clear consensus on how to define the solvating power of different electrolyte solvents \cite{chen2023correlating}.

Anion-rich solvation structures limit the interaction between free solvent molecules and \ce{Li+} ions, which helps suppress solvent decomposition and enhance electrolyte stability at both the anode and cathode interfaces \cite{li2023non}. These structures promote decomposition pathways dominated by the anion, leading to the formation of stable SEI layers enriched with inorganic components like LiF \cite{chen2020electrolyte}. As reported in the literature, these SEI layers play an important role in enhancing the performance of LIBs \cite{xu2014electrolytes, xu2004nonaqueous}. This enhancement is largely attributed to the SEI's ability to regulate \ce{Li+} ion migration at the electrode-electrolyte interface, which is governed by its composition and physicochemical properties \cite{xu2010differentiating, liu2019challenges}. Among its key features, the LiF-rich SEI layer functions as a robust protective shell on the electrode surface \cite{chen2020electrolyte}. Notably, an anion-rich solvation shell is known to facilitate the development of such LiF-rich SEI layers \cite{li2023non}. However, the strong coordination between \ce{Li+} and anions can raise the desolvation energy barrier, which may hinder \ce{Li+} transport—particularly in electrolytes with pronounced ion aggregation, such as those based on DPE or mixture of tetrahydrofuran (mixTHF) \cite{li2023non, chen2020electrolyte}. Prior research has demonstrated that DPE/\ce{LiFSI}-based electrolytes yield SEI layers with a high fluorine content ($\sim$43\%) and a significantly greater proportion of fluorinated species ($\sim$22\%) compared to other non-fluorinated ether systems like diethyl ether (DEE), 1,2-dimethoxyethane (DME), and diglyme (DIG) \cite{li2023non}. Furthermore, DPE/\ce{LiFSI} electrolytes are associated with enhanced cycling stability in lithium metal batteries \cite{li2023non}. These findings highlight the role of anion-driven interphase chemistry. Several studies have also focused on designing fluorinated electrolyte systems to  facilitate the formation of weakly solvated structures and LiF-rich SEI layers \cite{wang2021ion, zhang2023all, zou2023high, zhao2023electrolyte, piao2023stable, zhou2022integrated}. Based on this insight, we have designed a novel fluorinated FEME ether-based electrolyte featuring an anion-abundant solvation structure to promote the formation of these desirable SEI layers.

In this study, we computationally investigate \ce{LiFSI}-based non-aqueous electrolytes in fluorinated and non-fluorinated ether solvents, alongside highly soluble \ce{LiPF6}-based non-aqueous electrolytes in mixed carbonate solvents \cite{chen2024lino3}. The investigation includes a series of solvents: DPE, FEME, EC, and DEC. For ether-based electrolytes, a broad range of salt concentrations (1 M, 1.8 M, and 4 M) is selected, while for the mixed carbonate electrolyte, only 1 M is studied. Our results demonstrate that DPE- and FEME-based electrolytes exhibit anion-rich solvation structures even at a low salt concentration of 1 M, with the results for the DPE electrolyte aligning with previous studies \cite{li2023non}. In contrast, the mixed carbonate electrolyte predominantly features homogeneously dispersed SSIPs \cite{gullbrekken2024effect}. We begin by analyzing the electronic properties of the salt and solvent molecules, including HOMO/LUMO distributions and electrostatic potential (ESP) maps, to evaluate their chemical stability within the electrolyte. Next, fluorinated, non-fluorinated, and mixed carbonate-based electrolyte systems are modeled using molecular dynamics (MD) simulations to derive key properties of interest. Radial distribution functions (RDFs) and coordination numbers (CNs) are then calculated to provide insight into the solvation structure of these systems. The solvation structures of lithium ions are examined in detail, with a focus on the composition of the primary solvation shell, including the number of solvent molecules and anions. Additionally, the presence of solvent-separated ion pairs (SSIPs), contact ion pairs (CIPs), and aggregated species (AGGs) is analyzed for each electrolyte. Electronic characteristics of \ce{Li+} solvation structures are further explored using charge density difference analysis, Bader charge calculations, electrostatic potential maps, binding energies, and HOMO/LUMO distributions derived from density functional theory (DFT) calculations. This comprehensive approach provides valuable insights into the solvation behavior of fluorinated, non-fluorinated ether, and mixed carbonate-based electrolytes, contributing to the design of advanced electrolytes for lithium-ion battery applications.

\section{Computational Methods and Details}

\subsection{MD Simulations}

\begin{figure*}[h]
 \centering
 \includegraphics[width=\textwidth]{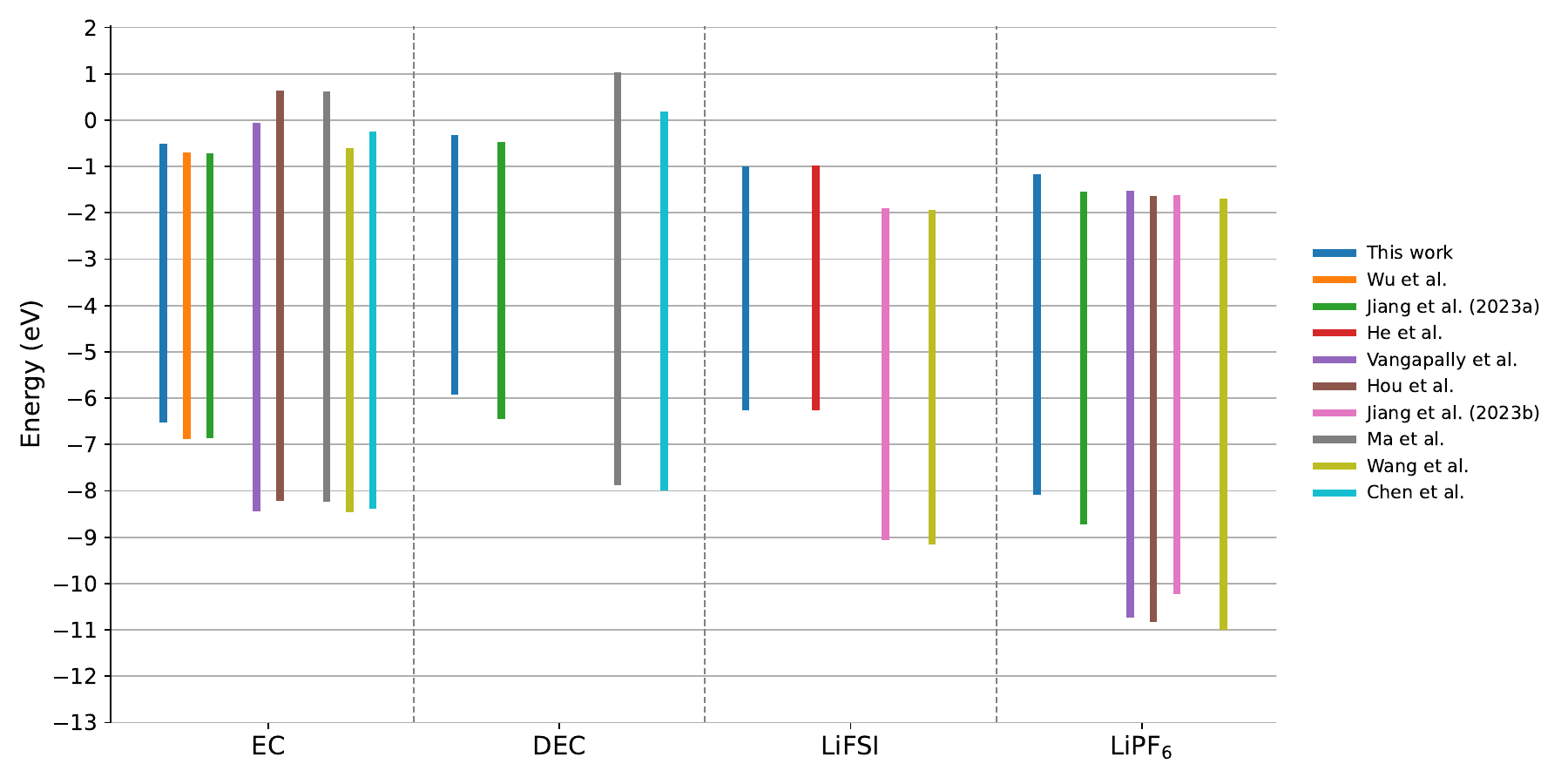}
 \caption{DFT validation of HOMO/LUMO energy levels of the studied salts and solvents. This work, Wu et al. \cite{wu2021synergistic}, Jiang et al. (2023a) \cite{jiang2023synergistic}, and He et al. \cite{he2022understanding} used the PBE functional in VASP, while Vangapally et al. \cite{vangapally2024fluorinated}, Hou et al. \cite{hou2025crosslinked}, Jiang et al. (2023b) \cite{jiang2023anion}, Ma et al. \cite{ma2023designing}, Wang et al. \cite{wang2020interface}, and Chen et al. \cite{chen2024regulate} used the B3LYP functional in Gaussian software.}
 \label{fig:homo_lumo_energylevel_compare}
\end{figure*}

In this study, non-fluorinated ether, fluorinated ether, and mixed organic carbonate solvents were considered. The selected solvents were dipropyl ether (DPE), 2,2-difluoroethyl methyl ether (FEME), ethylene carbonate (EC), and diethyl carbonate (DEC). Herein, seven types of electrolytes were studied: DPE+1 M \ce{LiFSI}, DPE+1.8 M \ce{LiFSI}, DPE+4 M \ce{LiFSI}, FEME+1 M \ce{LiFSI}, FEME+1.8 M \ce{LiFSI}, FEME+4 M \ce{LiFSI}, and 1:1 vol\% EC/DEC+1 M \ce{LiPF6}. The \ce{Li+} solvation structures in these electrolytes were investigated using classical molecular dynamics (MD) and density functional theory (DFT) simulations.

\begin{table}[h]
\small
  \caption{\ Density and molar mass of solvent molecules }
  \label{tbl:molecules}
  \begin{tabular*}{0.48\textwidth}{@{\extracolsep{\fill}}llll}
    \hline
    Solvent  & Density $(g/m^3)$ & Molar Mass $(g/mol)$ & \#Atoms \\
    \hline
    DPE & $736 \times 10^3$ & 102.177 & 21 \\
    FEME & $1004 \times 10^3$ & 96.076 & 12 \\
    EC & $132 \times 10^4$ & 88.06 & 10 \\
    DEC & $975 \times 10^3$ & 118.13 & 18 \\
    \hline
  \end{tabular*}
\end{table}

In each simulation box, lithium cations \ce{Li+} and \ce{FSI-} anions were randomly distributed among 542 solvent molecules for the DPE+1 M \ce{LiFSI}, DPE+1.8 M \ce{LiFSI}, and DPE+4 M \ce{LiFSI} systems. For the FEME+1 M \ce{LiFSI}, FEME+1.8 M \ce{LiFSI}, and FEME+4 M \ce{LiFSI} systems, the ions were randomly placed among 787 solvent molecules. Similarly, the 1:1 vol\% EC/DEC+1 M \ce{LiPF6} system contained 875 solvent molecules with \ce{Li+} and \ce{PF6-} ions arranged in a non-uniform manner. Table \ref{tbl:molecules} shows the density and molar mass of the solvent molecules \cite{ li2023single, moon2022non}. In Table \ref{tbl:electrolyte1}, the number of salt and solvent molecules in the electrolytes for any specific concentration was calculated (Supplementary Note 1\dag).



The initial configurations of these seven electrolyte systems were modeled using the PACKMOL \cite{martinez2009packmol} package by randomly placing the solvent molecules, \ce{FSI-}, \ce{PF6-}, and \ce{Li+} in a $5 \times 5 \times 5$ $\text{nm}^3$ cubic simulation box. Atomic and ionic interactions in the DPE-LiFSI, FEME-LiFSI, and EC-DEC-\ce{LiPF6} systems were described by the OPLS-AA (Optimized Potentials for Liquid Simulations All Atom) \cite{jorgensen1984optimized, jorgensen1996development} and ionic liquids force field. The bonded and non-bonded parameters of the OPLS-AA force field for the DPE, FEME, EC, and DEC solvent molecule atoms were obtained from LigParGen \cite{jorgensen2005potential, dodda20171, dodda2017ligpargen}, while the force field parameters for \ce{FSI-}, \ce{PF6-}, and \ce{Li+} ions were obtained from a database of several ionic liquids \cite{canongia2004modeling, shimizu2010molecular}. This OPLS-AA force field has been extensively validated for modeling lithium-ion battery electrolytes and offers a favorable balance between accuracy and computational efficiency.

The functional form of the OPLS force field is defined by a set of potential functions \cite{hou2021solvation} in Equations \ref{Eq:opls1} to \ref{Eq:opls7}, which include:

\begin{align}
\label{Eq:opls1}
E_{\text{total}}{(r^N)} &= E_{\text{bonded}} + E_{\text{nonbonded}} \\
\label{Eq:opls2}
E_{\text{bonded}} &= E_{\text{bonds}} + E_{\text{angles}} + E_{\text{dihedrals}} + E_{\text{impropers}} \\
\label{Eq:opls3}
E_{\text{bonds}} &= \sum_{\text{bonds}} K_r (r - r_0)^2 \\
\label{Eq:opls4}
E_{\text{angles}} &= \sum_{\text{angles}} K_\theta (\theta - \theta_0)^2 \\
\label{Eq:opls5}
E_{\text{dihedrals}} &= \sum_{\text{dihedrals}} V \left[ 1 + \cos(n\varphi - d) \right] \\
\label{Eq:opls6}
E_{\text{impropers}} &= \sum_{\text{impropers}} V \left[ 1 + d \cos(n\varphi) \right] \\
\label{Eq:opls7}
E_{\text{nonbonded}} &= \sum_{i>j} 4 \epsilon_{ij} \left[ \left(\frac{\sigma_{ij}}{r_{ij}}\right)^{12} - \left(\frac{\sigma_{ij}}{r_{ij}}\right)^6 \right] + \sum_{i>j} \frac{C q_i q_j}{\epsilon r_{ij}}
\end{align}

Where,
\[
\epsilon_{ij} = \sqrt{\epsilon_{ii} \epsilon_{jj}} \quad \text{and} \quad \sigma_{ij} = \frac{\sigma_{ii} + \sigma_{jj}}{2}
\]

In MD simulations, interactions within molecules are divided into bonded and non-bonded interactions. Bonded interactions, which include bonds, angles, dihedrals, and impropers, are modeled using harmonic functions. Non-bonded interactions include van der Waals forces and Coulombic forces, which describe the behavior between atoms that are not directly bonded. The dihedral term captures the torsional motion of four consecutively bonded atoms, and the improper term describes the torsional motion of three atoms arranged around a central fourth atom.

All MD simulations were performed using the LAMMPS \cite{thompson2022lammps} \href{https://lammps.org}{https://lammps.org} open-source software (version 23 Jun 2022). Lennard-Jones and Coulombic force interactions were cut off at a distance of 1.2 $\text{nm}$. Coulombic forces beyond the cutoff were computed using the particle-particle particle-mesh (PPPM) \cite{hockney2021computer} method to account for long-range electrostatic interactions with a relative error in forces of \(1 \times 10^{-5}\). Periodicity was applied in all the x, y, and z dimensions of the cubic simulation box. The equilibration procedure and production run are outlined as follows \cite{li2023non}. First, the prepared systems from the PACKMOL software were minimized using the steepest descent (SD) method with a convergence criterion of 1000 $\text{kcal/mol} \cdot \text{\AA}$, followed by conjugate gradient (CG) minimization with a convergence criterion of 10 $\text{kcal/mol} \cdot \text{\AA}$. Minimization algorithms were used to reduce the system's energy and prevent particle overlap. The systems were then equilibrated at a temperature of 298.15 K and a pressure of 1 atm in the isobaric-isothermal (NPT) ensemble using a time step of 1 fs for 2 ns to stabilize the potential energy and density of the systems. During equilibration, bond constraints were applied to specified bond lengths in the simulation using the SHAKE algorithm \cite{ryckaert1977numerical}. The temperature and pressure were regulated by the Nosé-Hoover thermostat and barostat \cite{shinoda2004rapid, hoover1985canonical, nose1984molecular}, with time constants set to produce characteristic fluctuations over 100 and 1000 time steps, respectively. Next, the equilibrated systems were heated to 500.15 K for 2 ns and then gradually cooled to 298.15 K over four steps, spanning 3 ns. Finally, production runs were performed in the canonical (NVT) ensemble at 298.15 K for 5 ns using a time step of 1 fs, from which the properties of interest were derived. The Nosé-Hoover thermostat was used in the NVT ensemble. All MD simulations and DFT calculations were carried out using our HPC cluster Wulver at NJIT and the Expanse supercomputing cluster at SDSC. 

\begin{table}[h]
\small
  \caption{\ Number of salt and solvent molecules in each electrolyte }
  \label{tbl:electrolyte1}
  \begin{tabular*}{0.48\textwidth}{@{\extracolsep{\fill}}llllll}
    \hline
    Electrolyte  & Concentration & \makecell{\#DPE \\ \#FEME \\ \#EC} & \#DEC & \makecell{\#\ce{LiFSI} \\ \#\ce{LiPF6}} & \#Atoms \\
    \hline
    DPE & 1 M & 542 & - & 75 & 12132\\
    DPE & 1.8 M & 542 & - & 135 & 12732\\
    DPE & 4 M & 542 & - & 301 & 14392\\
    FEME & 1 M & 787 & - & 75 & 10194\\
    FEME & 1.8 M & 787 & - & 135 & 10794\\
    FEME & 4 M & 787 & - & 301 & 12454\\
    1:1 EC/DEC & 1 M & 564 & 311 & 75 & 11838\\
    \hline
  \end{tabular*}
\end{table}

\begin{figure*}[ht!]
\centering
\begin{minipage}{.5\textwidth}
  \centering
  \subcaption*{a}
  \includegraphics[width=\linewidth]{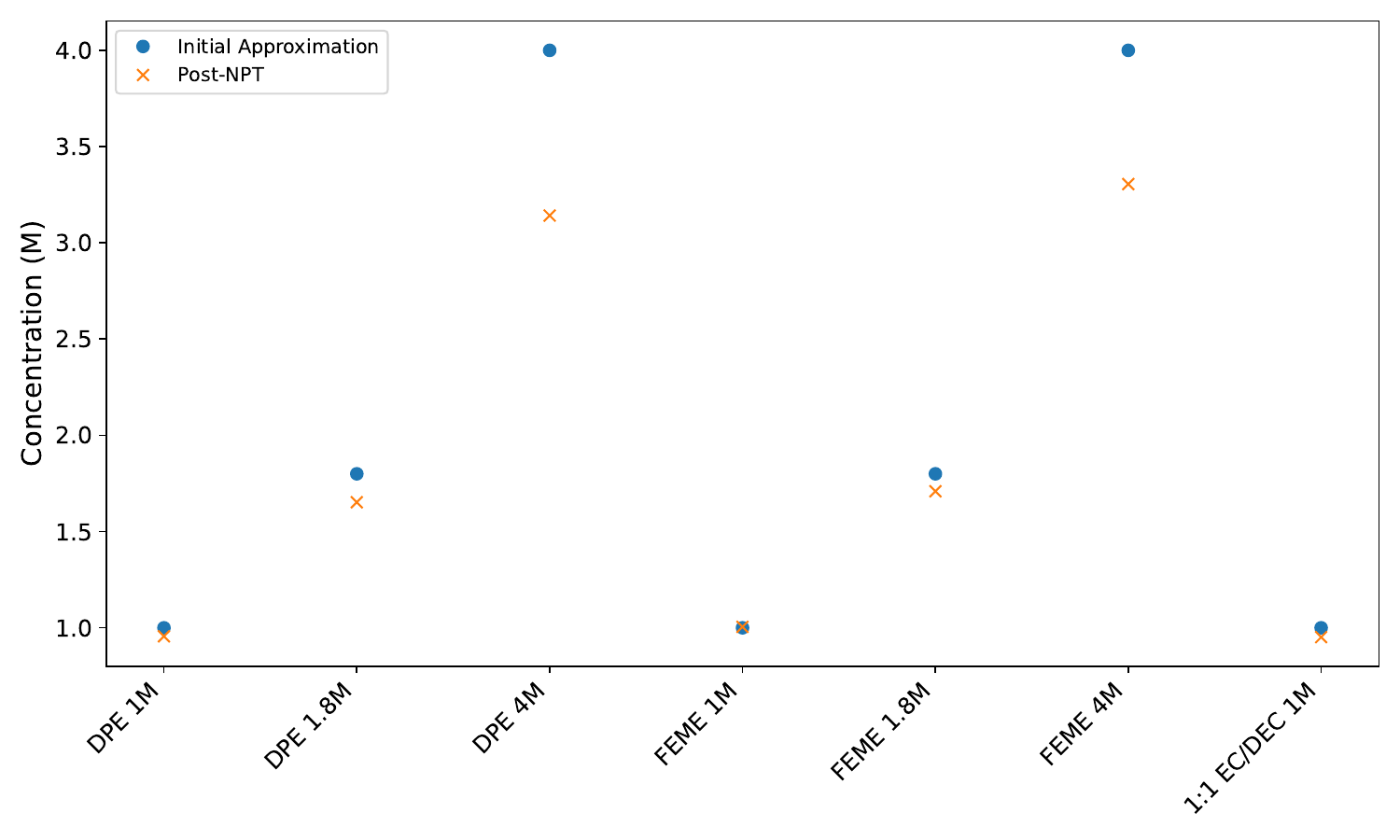}
  \phantomcaption
  \label{fig:postNPT_concentration_DPE_FEME}
\end{minipage}%
\hfill
\begin{minipage}{.5\textwidth}
  \centering
  \subcaption*{b}
  \includegraphics[width=\linewidth]{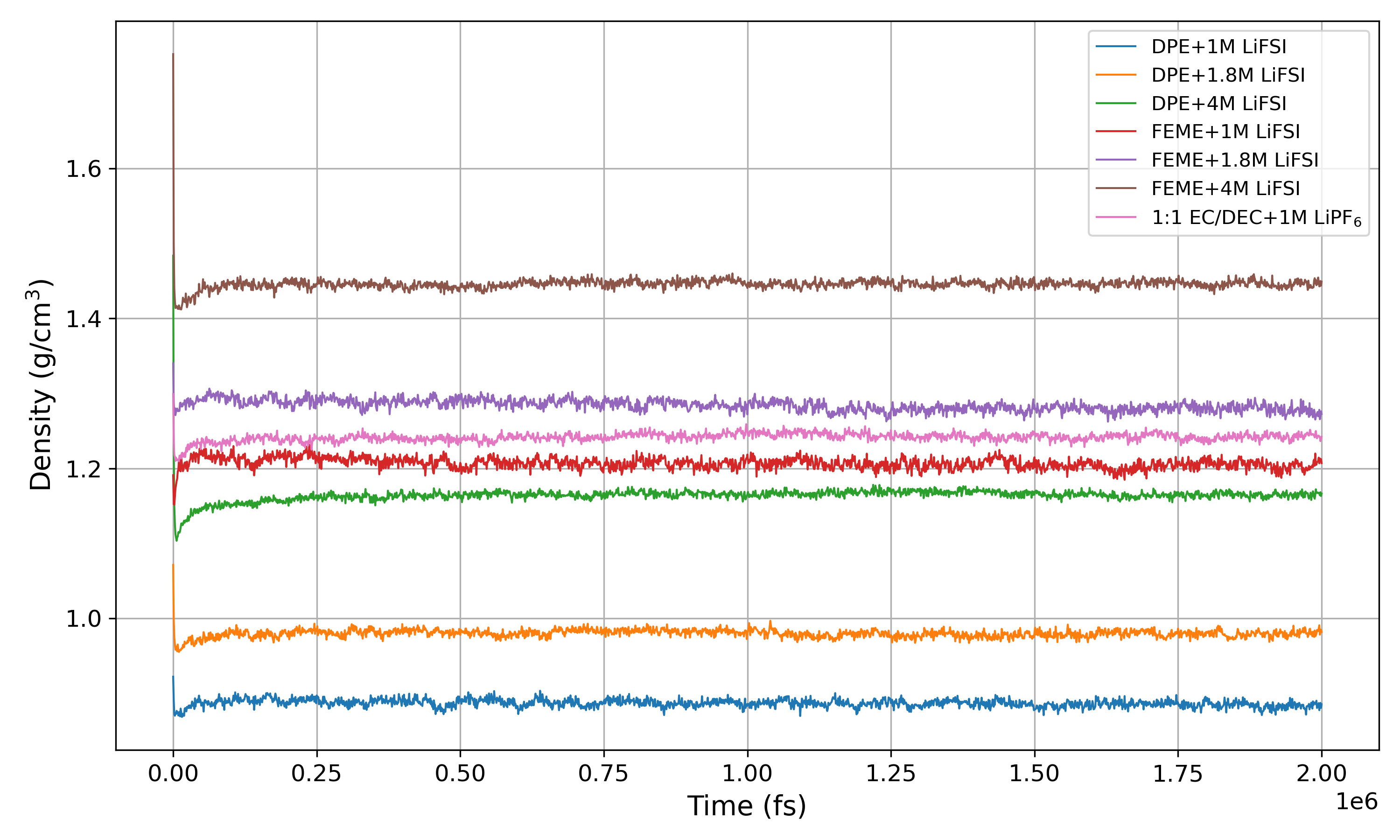}
  \phantomcaption
  \label{fig:Density_vs_Time}
\end{minipage}
\caption{(a) Initial approximate concentration before NPT equilibration and post-NPT concentration of the equilibrated system. (b) Density of the system during NPT equilibration.}
\label{fig:conc_density}
\end{figure*}

\begin{figure}[h]
 \centering
 \includegraphics[width=0.48\textwidth]{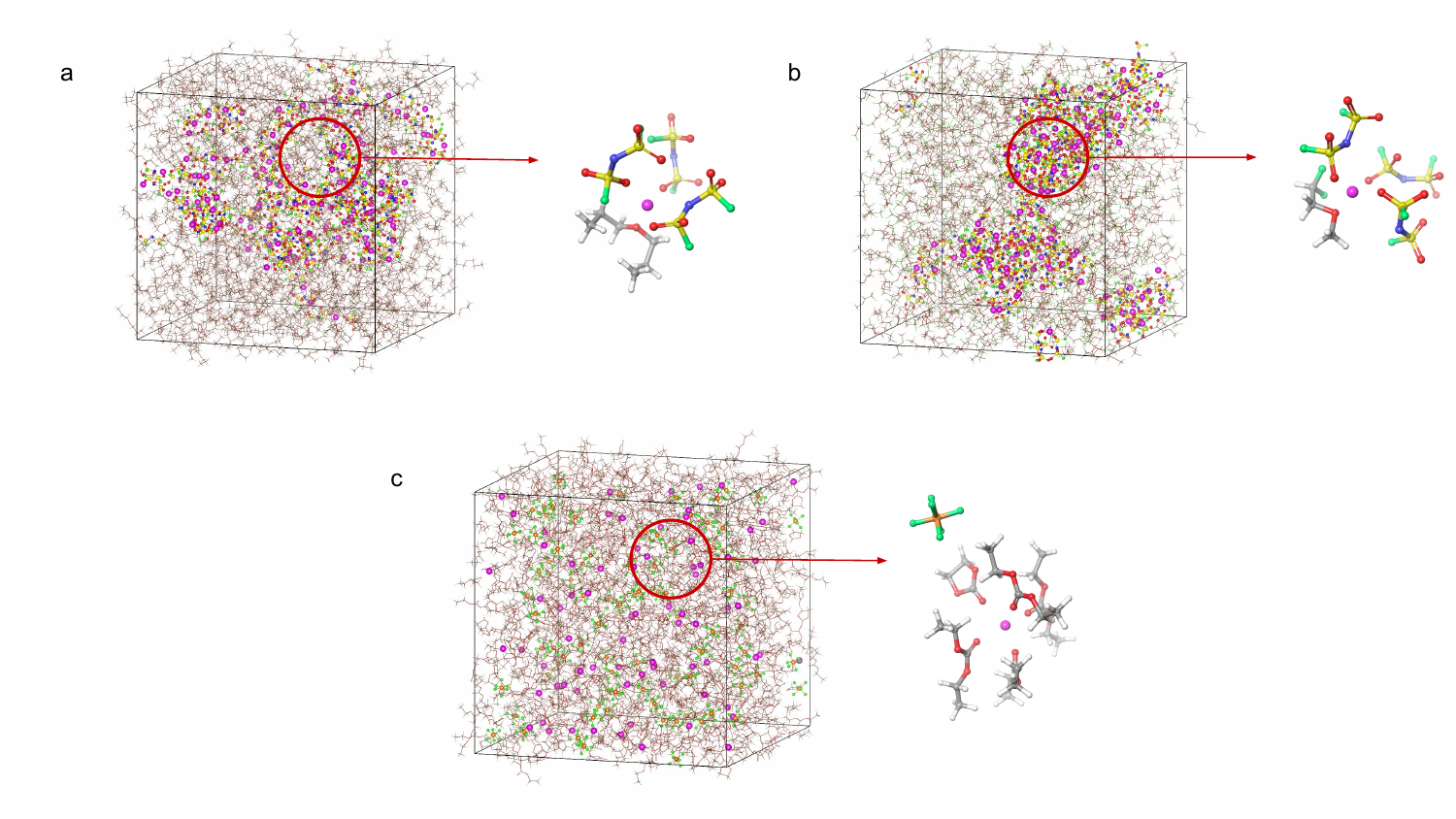}
 \caption{Snapshots of the simulation box obtained from MD simulations at 5 ns of the production run: (a) DPE+1.8 M \ce{LiFSI}, (b) FEME+1.8 M \ce{LiFSI}, and (c) 1:1 EC/DEC+1 M \ce{LiPF6}.}
 \label{fig:3D_box}
\end{figure}

\subsection{DFT Calculations}

In this research, all first-principles calculations in DFT were performed using the Vienna Ab Initio Simulation Package (VASP) software \cite{PhysRevB.54.11169, harl2010assessing}. The core-valence electron interactions were treated using the projector augmented wave method or PAW pseudopotentials \cite{blochl1994projector, kresse1999ultrasoft}. The commonly used Perdew-Burke-Ernzerhof (PBE) functional \cite{perdew1996generalized} under the Generalized Gradient Approximation (GGA) was used to model exchange and correlation interactions between electrons.

To perform geometry optimization in VASP, the atomic coordinates were allowed to change while keeping the shape and volume of the cell constant. Gaussian smearing was employed, with the smearing value set to 0.05 eV. The self-consistent field energy convergence was set to \(1 \times 10^{-6}\) eV, and the ionic force convergence tolerance was set to 0.02 eV/Å. An energy cutoff for the plane-wave basis set was specified at 520 eV. The Brillouin zone of the supercell was sampled using $\Gamma$-centered k-point grids (KPOINTS) of \(3 \times 3 \times 3\) with a k-mesh density of 0.03. Since the PBE functional provides a poor description of dispersion forces, the zero-damping DFT-D3 method of Grimme \cite{grimme2010consistent} was implemented to more accurately calculate the energy of the system. For calculations involving individual molecules, the size of the periodic models was set to \(10 \times 10 \times 10\) Å\(^3\), while for clusters, the size was set to \(15 \times 15 \times 15\) Å\(^3\) and \(20 \times 20 \times 20\) Å\(^3\). This setup ensures sufficient vacuum distance without significantly increasing the computational cost. In this study, all DFT calculations were conducted in vacuum. Both the geometry optimization and single point energy calculation, also called the self-consistent field (SCF) calculation were performed using non-spin-polarized calculations. The electronic structure information was obtained from SCF calculations performed on the optimized structures. This includes the ESP map, HOMO/LUMO distribution, Bader charge analysis, and charge density difference (CDD) of all the studied molecules and solvation structures \cite{zhang2021blue, kou2022effects, li2023non, li2024cation, zhao2022fluorinated, he2022understanding, wang2024weakly, zhou2022theoretical}. The binding energies were also calculated using DFT in vacuum with cluster models \cite{li2023non, atwi2022mispr, yang2023first}. For structure visualization, VESTA \cite{momma2011vesta} and Maestro were utilized, while the VASPKIT \cite{wang2021vaspkit} package was employed for post-processing the wave functions (WAVECAR) to generate the HOMO and LUMO distributions. The ESP maps, HOMO/LUMO diagrams, and CDD were also plotted using VESTA software.

\begin{figure*}[ht!]
\centering
\begin{minipage}{.51\textwidth}
  \centering
  \subcaption*{a}
  \includegraphics[width=\linewidth]{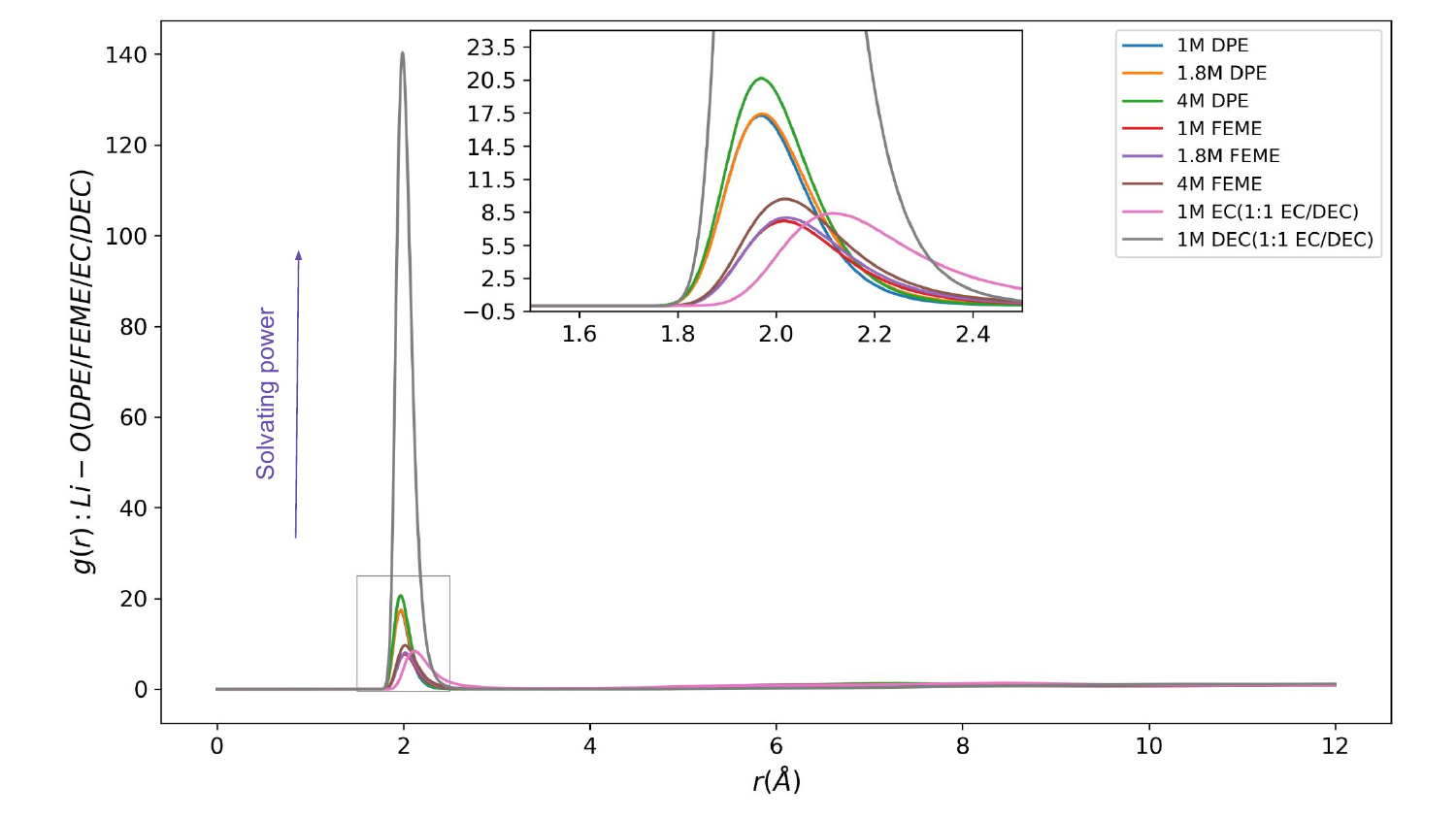}
  \phantomcaption
  \label{fig:RDF_Li_O(DPE/FEME/EC/DEC)}
\end{minipage}%
\hfill
\begin{minipage}{.48\textwidth}
  \centering
  \subcaption*{b}
  \includegraphics[width=\linewidth]{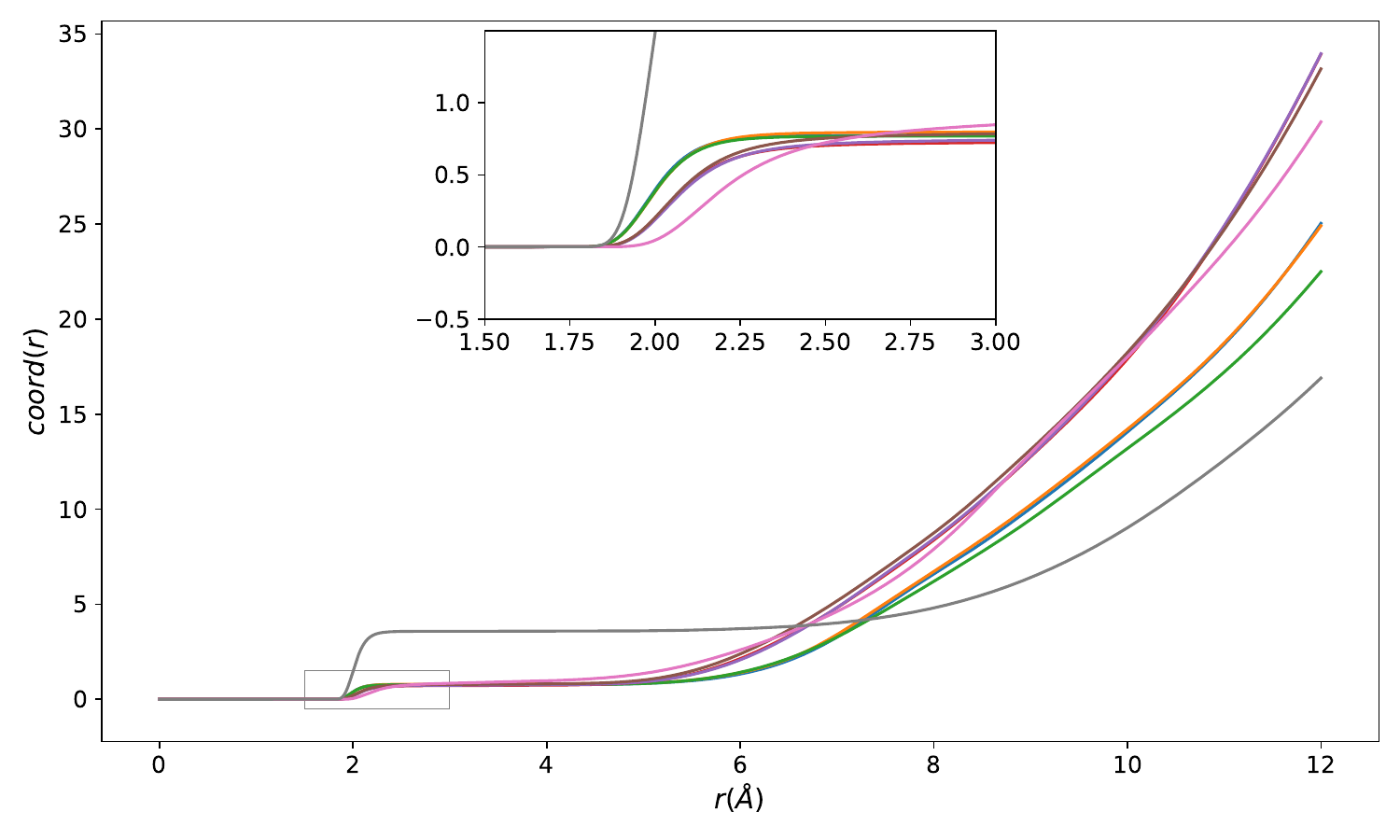}
  \phantomcaption
  \label{fig:CN_Li_O(DPE/FEME/EC/DEC)}
\end{minipage}

\vspace{-0.5em}

\begin{minipage}{.51\textwidth}
  \centering
  \subcaption*{c}
  \includegraphics[width=\linewidth]{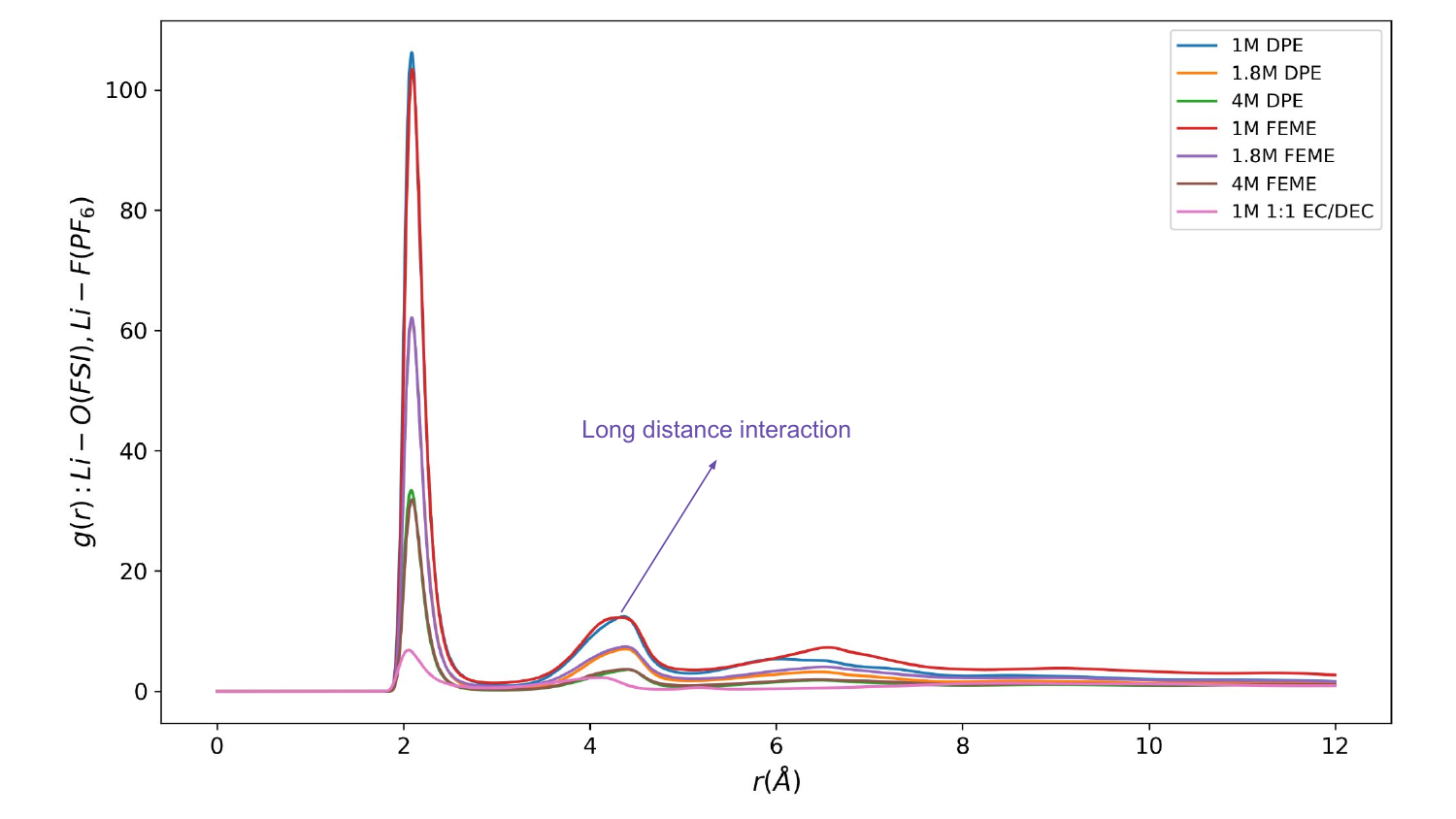}
  \phantomcaption
  \label{fig:RDF_Li_FSI/F(DPE/FEME/EC/DEC)}
\end{minipage}%
\hfill
\begin{minipage}{.48\textwidth}
  \centering
  \subcaption*{d}
  \includegraphics[width=\linewidth]{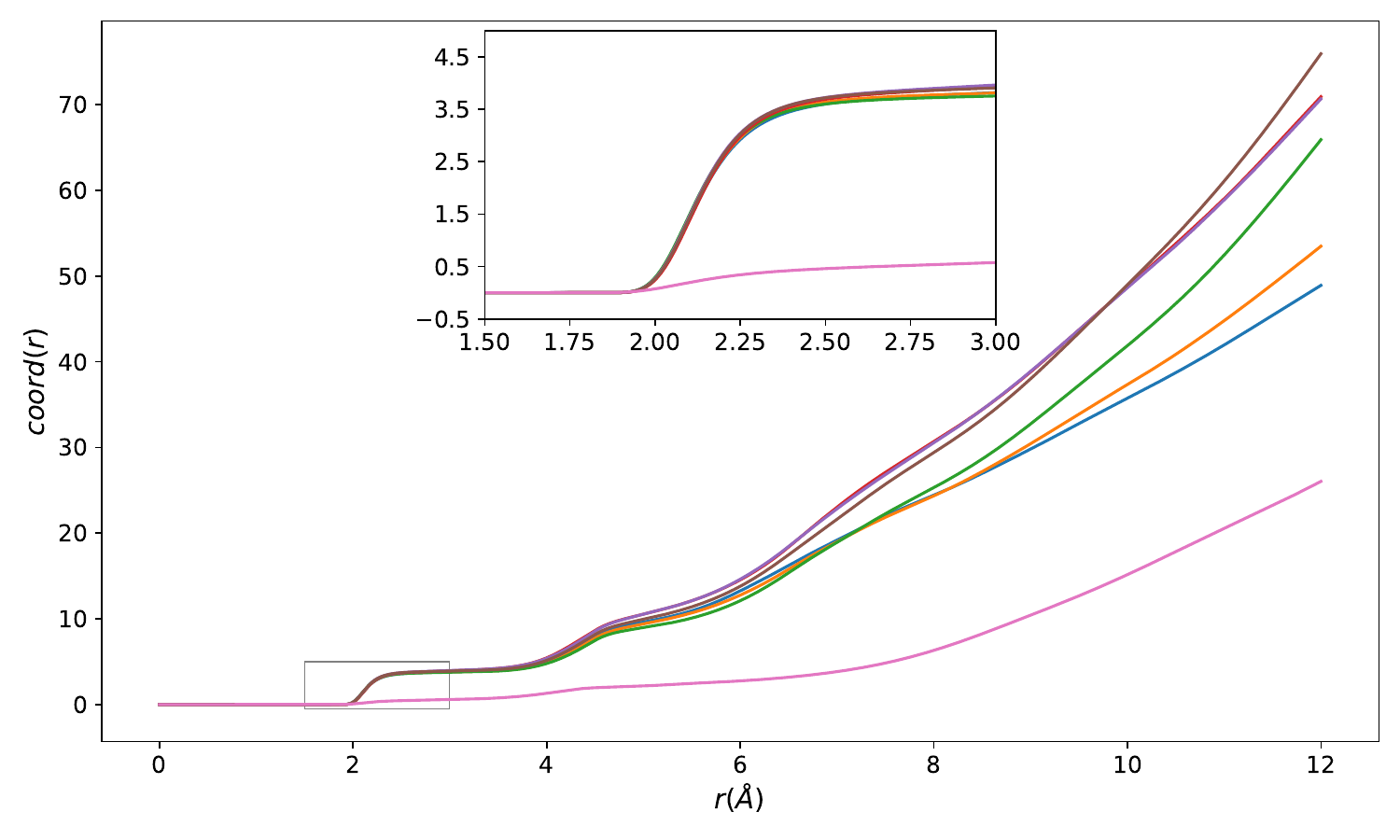}
  \phantomcaption
  \label{fig:CN_Li_FSI/F(DPE/FEME/EC/DEC)}
\end{minipage}

\caption{(a) RDFs of \ce{Li+-O(DPE/FEME/EC/DEC)} and corresponding (b) coordination numbers as a function of distance. (c) RDFs of \ce{Li+-O(FSI-}), \ce{Li+-F(PF6-}), and corresponding (d) coordination numbers as a function of distance.}
\label{fig:RDF_CN_Li_O/FSI/F}
\end{figure*}

\section{Results and Discussion}

\subsection{Salts, Solvents, and Electrolyte Systems}

A variety of solvents, including non-fluorinated ether (DPE), fluorinated ether (FEME), and carbonate solvents (EC and DEC) were selected to investigate the \ce{Li+} solvation structures and coordination in the electrolyte (Fig. \ref{fig:compact_figures_ESP} and Table \ref{tbl:molecules}). All MD production runs were conducted at 25°C. The ether-based electrolytes used \ce{LiFSI}, while the mixed carbonate-based electrolyte was prepared with \ce{LiPF6} salt. To systematically compare solvation structures across different solvent environments, we selected \ce{LiFSI} for both DPE and FEME ether-based electrolytes due to its tendency to promote anion-rich solvation structures even at relatively low concentrations. The combination of fluorinated FEME with \ce{LiFSI} contributes to the formation of a robust LiF-rich SEI layer, which is critical for interfacial stability \cite{chen2020electrolyte}. Prior studies have also reported anion-rich solvation in similar WSEE systems, such as DPE+1.8 M \ce{LiFSI}, DEE+1.8 M \ce{LiFSI}, and FDMB+1 M \ce{LiFSI} \cite{li2023non, yu2020molecular, wang2024ether}. In contrast, \ce{LiPF6} was selected for the carbonate-based EC/DEC system due to its widespread use in commercial LIBs and its formation of SSIPs \cite{gullbrekken2024effect}. This makes it a meaningful benchmark for comparison with our ether-based systems. Using \ce{LiFSI} salt allowed us to compare the anion-rich solvation structure of the FEME electrolyte with the SSIP-dominated carbonate-based electrolyte (\ce{LiPF6} in EC/DEC), while also enabling a controlled comparison with non-fluorinated DPE to highlight the enhanced ion aggregation promoted by the FEME+\ce{LiFSI} system.

DFT calculations in VASP were used to simulate the electrostatic potential (ESP) maps, Highest Occupied Molecular Orbital (HOMO), and Lowest Unoccupied Molecular Orbital (LUMO) of the salt/solvent molecules. Fig. \ref{fig:compact_figures_ESP} shows regions of high electron density (negative charge) and low electron density (positive charge) in the ESP maps, with values ranging from 0 to 1.24 $e/bohr^3$ (DPE, EC, DEC) and from 0 to 2.08 $e/bohr^3$ (FEME, \ce{LiFSI}, \ce{LiPF6}). The isosurfaces and energy levels of the HOMO/LUMO distributions for DPE, FEME, EC, DEC, \ce{LiFSI}, and \ce{LiPF6} molecules are depicted in Fig. \ref{fig:HOMOLUMO}. The higher the HOMO, the easier it is for the molecule to donate electrons, while a lower LUMO indicates it can accept electrons more easily. The energy difference between the HOMO and LUMO is referred to as the energy band gap $(LUMO - HOMO)$. A narrower band gap usually corresponds to greater chemical reactivity and lower stability, while a wider band gap suggests less reactivity and greater stability. Herein, the lower HOMO (-6.2697 eV) and LUMO (-0.9929 eV) of \ce{LiFSI} compared to DPE and FEME indicate it will decompose first during the charge/discharge cycle \cite{he2022understanding}. Likewise, the lower HOMO (-8.0773 eV) and LUMO (-1.1581 eV) of \ce{LiPF6} relative to EC and DEC suggest \ce{LiPF6} will decompose before the solvents and earlier than \ce{LiFSI}. According to the literature, slight variations in the HOMO/LUMO energy values of the same molecule can result from using different calculation methods (PBE, GGA, B3LYP functional) and software packages such as VASP and Gaussian \cite{wu2021synergistic, jiang2023synergistic, he2022understanding, vangapally2024fluorinated, hou2025crosslinked, jiang2023anion, ma2023designing, wang2020interface, chen2024regulate, aoki2022effective, wu2024recent, bai2024perspective, guo2022inner, song2023boosting, guo2024poly, jin2023ethyl, yang2023tuning}. The DFT validations are shown in Fig. \ref{fig:homo_lumo_energylevel_compare}. Our HOMO/LUMO energy values of the salts and solvents are quite similar to those obtained using the PBE functional and VASP.

The electrolyte systems were modeled using MD simulations in LAMMPS. Each system followed a process of minimization, equilibration, melting, quenching, and a 5 ns production run, from which key properties such as radial distribution functions (RDF), coordination numbers (CN), and solvation structures were obtained. Fig. \ref{fig:3D_box} shows the 3D simulation box for the 1.8 M and 1 M DPE-\ce{LiFSI}, FEME-\ce{LiFSI}, and EC-DEC-\ce{LiPF6} electrolyte systems after the production run. The supplementary section includes OVITO-generated GIF showing the 5 ns production run trajectories for these electrolyte systems \cite{stukowski2009visualization}. The 3D simulation boxes of all the remaining systems, including DPE-\ce{LiFSI} and FEME-\ce{LiFSI} at 1 M and 4 M concentrations are depicted in Fig. S1\dag. Additionally, Fig. S2\dag, S3\dag, S4\dag, and S5\dag\ provide the volume, density, pressure, and temperature of all the equilibrated systems. During equilibration, the pressure and temperature fluctuated around 1 atm and 298.15 K, respectively. Larger systems with more atoms (4 M) exhibited fewer fluctuations compared to smaller systems (1 M and 1.8 M). The initial approximate concentration (before NPT equilibration), the actual concentration after NPT equilibration, and the system density during the NPT equilibration process are shown in Fig. \ref{fig:conc_density}. The density of our equilibrated system, 1 M \ce{LiPF6} in a 1:1 EC/DEC mixture is 1.240 $g/cm^3$ at 25°C. This value is in good agreement with experimental data reported in the literature: Anton Paar reports 1.242 $g/cm^3$ at 20°C, and Sigma-Aldrich lists 1.26 $g/cm^3$ at 25°C. Additionally, published values include 1.225 $g/cm^3$ at 25°C from Lundgren et al., 1.23 $g/cm^3$ from Lee et al., and 1.25 $g/cm^3$ from Dougassa et al., all at 25°C \cite{lundgren2014electrochemical, lee2002study, dougassa2013low, landry2025density}.

\begin{figure*}[ht!]
\centering
\begin{minipage}{.51\textwidth}
  \centering
  \subcaption*{a}
  \includegraphics[width=\linewidth]{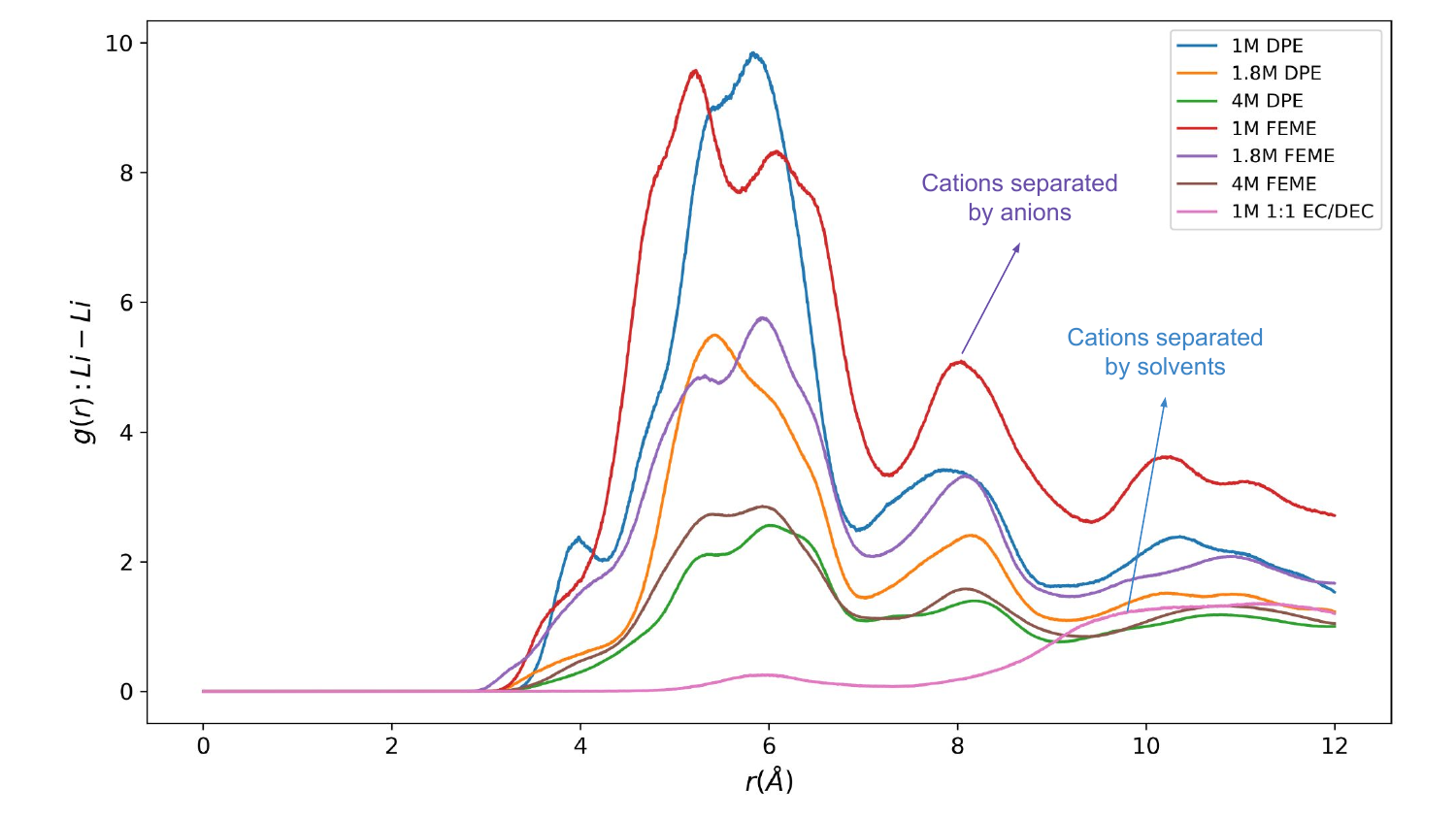}
  \phantomcaption
  \label{fig:RDF_Li_Li(DPE/FEME/EC/DEC)}
\end{minipage}%
\hfill
\begin{minipage}{.48\textwidth}
  \centering
  \subcaption*{b}
  \includegraphics[width=\linewidth]{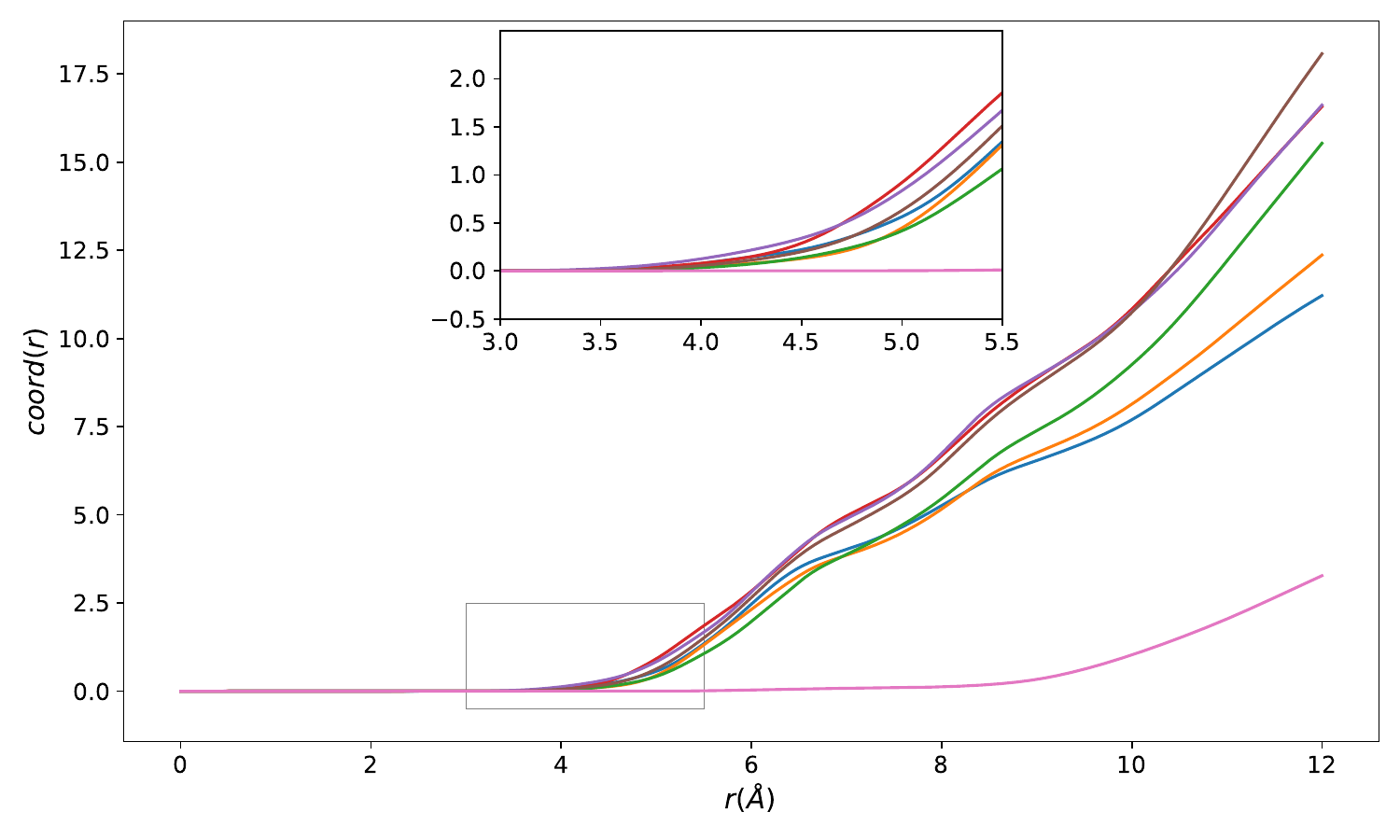}
  \phantomcaption
  \label{fig:CN_Li_Li(DPE/FEME/EC/DEC)}
\end{minipage}

\vspace{-0.5em}

\begin{minipage}{.48\textwidth}
  \centering
  \subcaption*{c}
  \includegraphics[width=\linewidth]{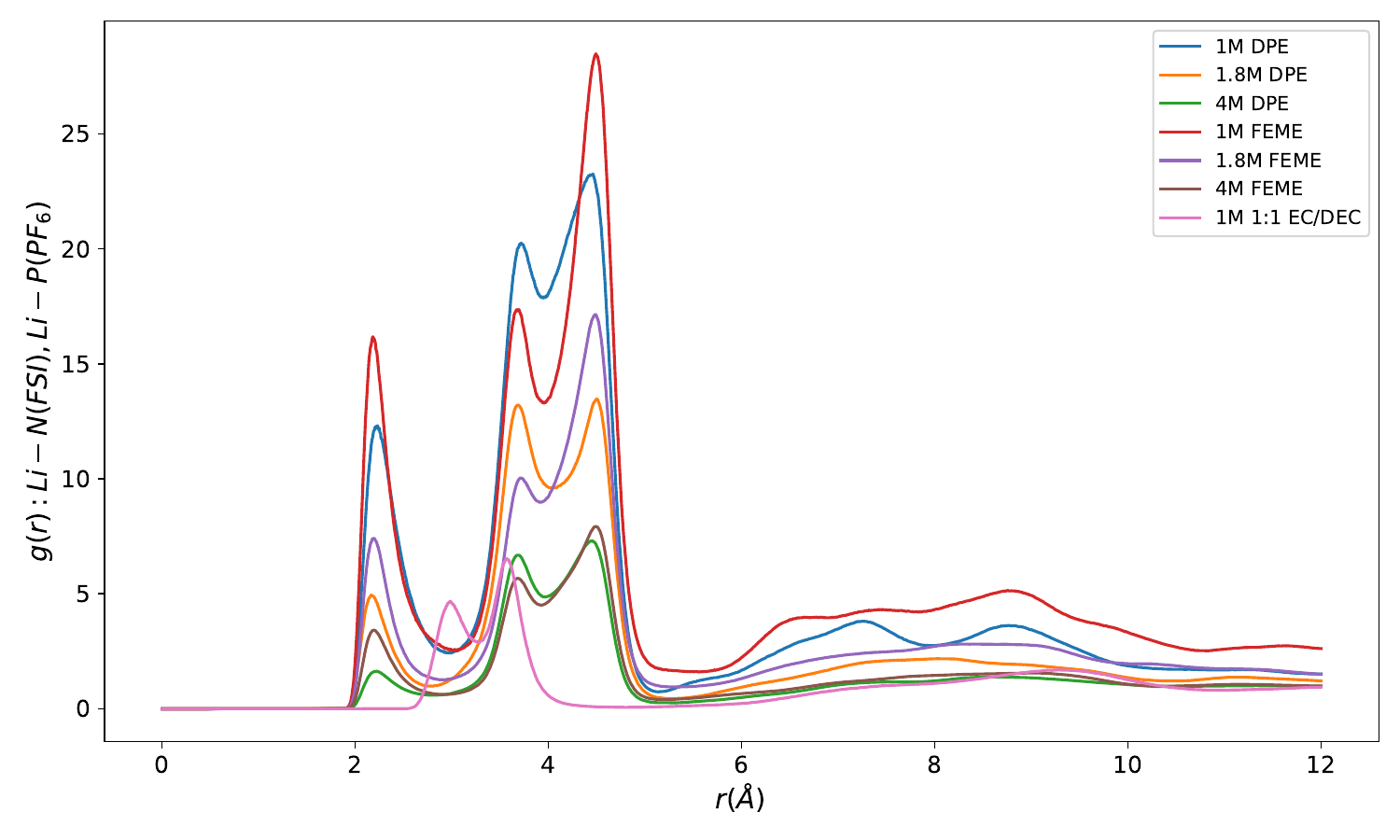}
  \phantomcaption
  \label{fig:RDF_Li_N/P(DPE/FEME/EC/DEC)}
\end{minipage}%
\hfill
\begin{minipage}{.48\textwidth}
  \centering
  \subcaption*{d}
  \includegraphics[width=\linewidth]{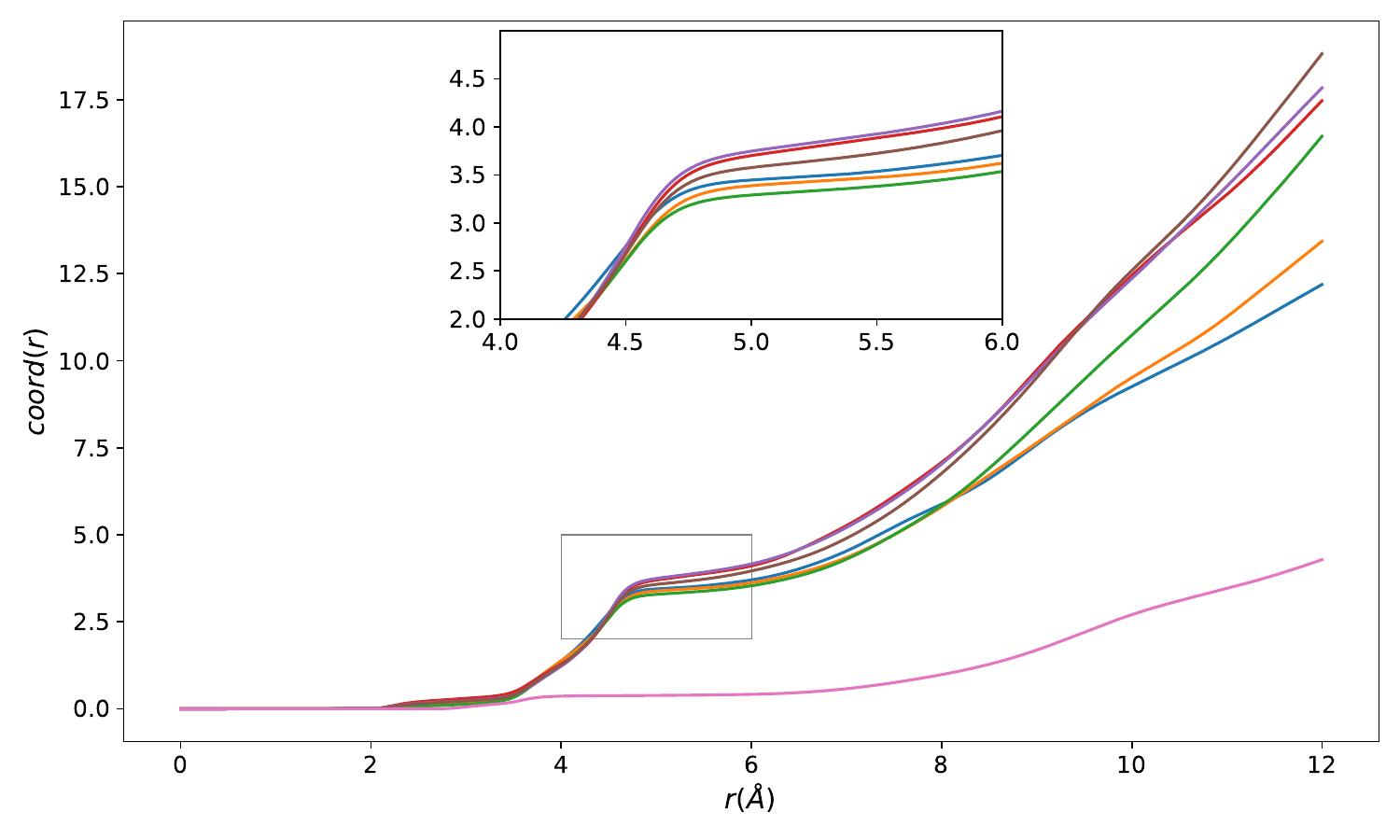}
  \phantomcaption
  \label{fig:CN_Li_N/P(DPE/FEME/EC/DEC)}
\end{minipage}
\caption{(a) RDFs of \ce{Li+-Li+} and corresponding (b) coordination numbers as a function of distance. (c) RDFs of \ce{Li+-N(FSI-}), \ce{Li+-P(PF6-}), and corresponding (d) coordination numbers as a function of distance.}
\label{fig:RDF_CN_Li_Li/N/P}
\end{figure*}

\subsection{Radial Distribution Function and Coordination Number}

The radial distribution function (RDF) and coordination number (CN) of the electrolytes were calculated during MD simulations to study the electrolyte structure and the coordination environment of \ce{Li+}. The RDF and CN for \ce{Li+-O(DPE)}, \ce{Li+-O(FEME)}, \ce{Li+-O(EC)}, \ce{Li+-O(DEC)}, \ce{Li+-O(FSI-}), \ce{Li+-N(FSI-}), \ce{Li+-P(PF6-}), \ce{Li+-F(PF6-}), and \ce{Li+-Li+} were calculated by averaging across 1 million configurations of each electrolyte system during the last 1 ns of the 5 ns production run. These results are shown in Fig. \ref{fig:RDF_CN_Li_O/FSI/F} and \ref{fig:RDF_CN_Li_Li/N/P} (Fig. S6\dag\ and S7\dag), with the corresponding numerical data provided in Table \ref{tbl:RDF_CN}. For more detailed information, Fig. S8\dag\ to S14\dag\ also provide the time-averaged RDF and CN plots for each 1 ns interval of the 5 ns production run. Since the RDF and CN remained consistent across each 1 ns interval, a total production run of 5 ns was selected. 

In our simulations, weak solvating power was identified in ether solvents (DPE and FEME), as indicated by their low RDF peaks, low coordination numbers, and weak binding energies with \ce{Li+}. Conversely, in the mixed carbonate systems, EC and DEC exhibited stronger coordination with \ce{Li+}, evident from sharper RDF peaks, higher CNs, and greater binding energies, suggesting stronger solvating power. This contrast highlights the impact of solvent type—ether versus carbonate—on solvation behavior and ion coordination. Such findings are crucial for tailoring electrolytes to achieve desired interfacial properties in LIBs. These results show that solvating power is not determined by a single parameter but arises from a combination of molecular properties and competitive ion interactions within the solvation shell \cite{chen2023correlating, su2019solvating, reichardt2021solvation}. In this work, two carbonate solvents in a mixed-solvent electrolyte, with a solvating power order of EC $<$ DEC (computationally), and two ether solvents in a single-solvent electrolyte, with a solvating power order of FEME $<$ DPE were investigated.

The RDF and CN were calculated using Equations \ref{Eq:RDF} and \ref{Eq:CN}, where $n(r)$ is the average number of particles in the spherical shell, $r$ is the interatomic separation distance, and $\rho$ is the particle density in the system. As reported in the literature, the cutoff or threshold value of the pairwise distance ($r$) for the RDF corresponds to the distance at the maximum peak, while for the CN, it is the first minimum following the first peak in the RDF, representing the first solvation shell \cite{gullbrekken2024effect}.

\begin{gather}
\label{Eq:RDF}
    g(r) = \frac{n(r)}{\rho 4\pi r^2 dr} \\
\label{Eq:CN}
    coord(r) = 4 \pi \rho \int_{0}^{r_{\text{c}}} g(r) r^2 dr
\end{gather}

In this study, three different salt concentrations (1 M, 1.8 M, and 4 M) were used for both DPE and FEME electrolytes, whereas only a 1 M concentration was applied to the EC/DEC electrolyte. The \ce{Li+-solvent} and \ce{Li+-anion} pairwise interactions follow similar trends in each ether electrolyte but exhibit opposite trends in the carbonate electrolyte (Fig. \ref{fig:RDF_CN_Li_O/FSI/F}, \ref{fig:RDF_CN_Li_Li/N/P}, and Table \ref{tbl:RDF_CN}). In the mixed EC/DEC electrolyte, DEC solvents exhibit stronger coordination with \ce{Li+} compared to EC, as indicated by the sharp RDF peaks of \ce{Li+-O(DEC)} (Fig. \ref{fig:RDF_Li_O(DPE/FEME/EC/DEC)} and \ref{fig:CN_Li_O(DPE/FEME/EC/DEC)}) \cite{ gullbrekken2024effect, seo2015role, zhang2023all, zou2023high}. Conversely, the minimal \ce{Li+-F(PF6-}) RDF peaks suggest that the \ce{PF6-} anion has little influence on \ce{Li+} solvation structures (Fig. \ref{fig:RDF_Li_FSI/F(DPE/FEME/EC/DEC)} and \ref{fig:CN_Li_FSI/F(DPE/FEME/EC/DEC)}). On the other hand, Fig. \ref{fig:RDF_CN_Li_O/FSI/F} shows that in both DPE and FEME electrolytes, due to their weak solvating power as indicated by the negligible RDF peaks of \ce{Li+-O(DPE)} and \ce{Li+-O(FEME)}, DPE and FEME solvents do not strongly interact with the cations. This allows \ce{Li+} to form strong interactions with the \ce{FSI-} anion in the primary solvation shell ($r$ $\approx$ 2.09 Å), as shown by the pronounced RDF peaks of \ce{Li+-O(FSI-}) \cite{li2023non, holoubek2021tailoring, li2024branch, bai2024perspective}. From Fig. \ref{fig:RDF_Li_O(DPE/FEME/EC/DEC)}, \ref{fig:RDF_Li_FSI/F(DPE/FEME/EC/DEC)}, and Table \ref{tbl:RDF_CN}, in both DPE and FEME electrolytes, as salt concentration increases, the maximum RDF peaks for \ce{Li+-O(DPE)} and \ce{Li+-O(FEME)} increase, while those for \ce{Li+-O(FSI-}) decrease. It is found that the maximum RDF peaks of \ce{Li+-O(FSI-}) are consistently higher than those of \ce{Li+-O(DPE)} and \ce{Li+-O(FEME)}, which further confirms the weak solvating power of DPE and FEME solvents across varying salt concentrations. Additionally, the lower RDF peaks and coordination number of \ce{Li+-O(FEME)} compared to \ce{Li+-O(DPE)} suggest that FEME has weaker solvating power than DPE. This observation is also supported by the coordination number of \ce{Li+-O(FSI-}), which is slightly higher in FEME (ranging from 3.91 to 3.96) than in DPE (ranging from 3.75 to 3.82) electrolytes (Table \ref{tbl:RDF_CN}). The cutoffs for the sharp RDF peaks of \ce{Li+-O(DPE)}, \ce{Li+-O(FEME)}, and \ce{Li+-O(FSI-}) are observed at $r$ = 1.97 Å, $r$ = 2.02 Å, and $r$ = 2.09 Å, respectively, and remain unchanged across different salt concentrations.

\begin{figure}[ht!]
 \centering
 \includegraphics[width=.48\textwidth]{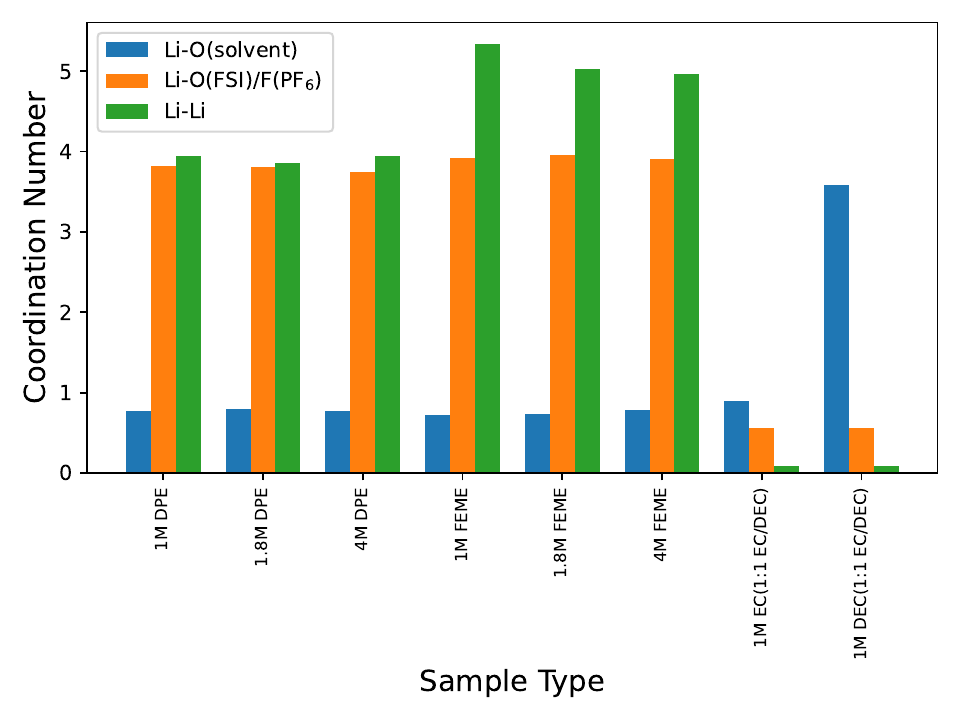}
 \caption{Atomic coordination numbers of the electrolytes. The CN of \ce{Li+-Li+} and \ce{Li+-F(PF6-}) in mixed EC/DEC carbonate electrolyte is assumed to be the same for both EC and DEC solvents.}
 \label{fig:CN_bar}
\end{figure}

From Fig. \ref{fig:RDF_Li_FSI/F(DPE/FEME/EC/DEC)}, \ref{fig:RDF_Li_Li(DPE/FEME/EC/DEC)}, and \ref{fig:RDF_Li_N/P(DPE/FEME/EC/DEC)}, the RDFs of \ce{Li+-O(FSI-}), \ce{Li+-F(PF6-}), \ce{Li+-N(FSI-}), \ce{Li+-P(PF6-}), and \ce{Li+-Li+} suggest the formation of ion clusters in all the studied electrolytes. Fig. \ref{fig:RDF_Li_Li(DPE/FEME/EC/DEC)} shows that shorter \ce{Li+-Li+} interaction distances are observed at $r$ = 5.83 Å, 5.43 Å, 6.00 Å (1st peak) and 7.85 Å, 8.15 Å, 8.17 Å (2nd peak) for the DPE electrolytes, and $r$ = 5.23 Å, 5.94 Å, 5.93 Å (1st peak) and 8.04 Å, 8.08 Å, 8.09 Å (2nd peak) for the FEME electrolytes, corresponding to 1 M, 1.8 M, and 4 M concentrations, respectively. Additionally, in these ether electrolytes, shorter \ce{Li+-N(FSI-}) interaction distances are also observed at $r$ $\approx$ 2.20 Å and $r$ $\approx$ 4 Å, where two consecutive RDF peaks are identified around 4 Å (Fig. \ref{fig:RDF_Li_N/P(DPE/FEME/EC/DEC)}). These shorter interaction distances of \ce{Li+-Li+} and \ce{Li+-N(FSI-}) indicate the presence of large ion aggregates (AGGs) composed of multiple \ce{Li+} and \ce{FSI-} ions \cite{cao2021effects, li2023non}. Across all the salt concentrations (1 M, 1.8 M, and 4 M) in DPE and FEME ether electrolytes, the \ce{Li+} and \ce{FSI-} ions form large ion aggregates through bridging coordination, where the \ce{FSI-} anions coordinate with multiple \ce{Li+} cations via their O atoms (Fig. \ref{fig:3D_box} and S1\dag). This aggregation behavior is further supported by the long-range \ce{Li+-O(FSI-}) interactions in the secondary solvation shell, observed at $r$ $\approx$ 4.38 Å in DPE and 4.36 Å in FEME electrolytes (Fig. \ref{fig:RDF_Li_FSI/F(DPE/FEME/EC/DEC)}). In contrast, the EC/DEC carbonate electrolyte exhibits a homogeneous distribution of \ce{Li+}, \ce{PF6-}, and EC/DEC solvent molecules (Fig. \ref{fig:3D_box}). The \ce{Li+} cations in the carbonate electrolyte are more widely separated as they are strongly coordinated by solvent molecules, which leads to the formation of solvent separated ion pairs (SSIPs) \cite{gullbrekken2024effect, skarmoutsos2015li+, koo2023role, aoki2022predictive, lee2023boosting, jia2020controlling, chen2020electrolyte}. This is confirmed by the lower RDF peaks and coordination numbers of \ce{Li+-Li+} interactions in the primary solvation shell (Fig. \ref{fig:RDF_Li_Li(DPE/FEME/EC/DEC)} and \ref{fig:CN_Li_Li(DPE/FEME/EC/DEC)}). Similarly, the lower RDF peaks and CN of \ce{Li+-F(PF6-}) and \ce{Li+-P(PF6-}) interactions also confirm the significant presence of SSIPs in this carbonate electrolyte (Fig. \ref{fig:RDF_Li_FSI/F(DPE/FEME/EC/DEC)}, \ref{fig:CN_Li_FSI/F(DPE/FEME/EC/DEC)}, \ref{fig:RDF_Li_N/P(DPE/FEME/EC/DEC)}, and \ref{fig:CN_Li_N/P(DPE/FEME/EC/DEC)}). In this study, for all seven electrolytes, the cutoff value used to calculate the coordination number of \ce{Li+-O(solvent)} and \ce{Li+-O(anion)} is approximately 3 Å (3.49 Å for \ce{Li+-O(EC)}), and for \ce{Li+-Li+}, it is approximately 7 Å (Table \ref{tbl:RDF_CN}). Within this cutoff, the CN of \ce{Li+-O(solvent)} is lower in FEME (0.72, 0.74, 0.79) compared to DPE (0.77, 0.80, 0.77), while the CN of \ce{Li+-O(FSI-}) is higher in FEME (3.92, 3.96, 3.91) than in DPE (3.82, 3.81, 3.75) at 1 M, 1.8 M, and 4 M concentrations, respectively. Similarly, the CN of \ce{Li+-Li+} interactions is higher in FEME (5.34, 5.03, 4.96) than in DPE (3.95, 3.86, 3.94), whereas it is significantly lower in the EC/DEC system (0.09). In all electrolytes, the RDF and CN of \ce{Li+-Li+} interactions remain zero up to $r$ = 3 Å (Fig. \ref{fig:RDF_Li_Li(DPE/FEME/EC/DEC)} and \ref{fig:CN_Li_Li(DPE/FEME/EC/DEC)}). The atomic coordination numbers for all electrolytes are also compared in Fig. \ref{fig:CN_bar}. Our findings reveal that DPE and FEME electrolytes are primarily composed of AGGs, while SSIPs are more prevalent in the EC/DEC electrolyte. Inside the first solvation shell, the higher RDF peaks and CN values of \ce{Li+-N(FSI-}) in FEME electrolytes also highlight that \ce{Li+-FSI-} pairs exhibit stronger aggregation in FEME electrolytes than in DPE electrolytes across all the salt concentrations (Fig. \ref{fig:3D_box}, \ref{fig:RDF_Li_N/P(DPE/FEME/EC/DEC)}, \ref{fig:CN_Li_N/P(DPE/FEME/EC/DEC)}, S1\dag, and Table \ref{tbl:RDF_CN}). This is further supported by the higher coordination numbers of \ce{Li+-Li+} and \ce{Li+-O(FSI-}) interactions in the primary solvation shell of FEME electrolytes compared to all other electrolytes (Table \ref{tbl:RDF_CN}). Our RDF results and solvation structures for DPE+1.8 M \ce{LiFSI} using the OPLS-AA \cite{jorgensen1984optimized, jorgensen1996development} force field are quite similar to those reported by Li et al. using the OPLS-2005 force field from Schrödinger \cite{li2023non, banks2005integrated}. The properties of the 1:1 EC/DEC+1 M \ce{LiPF6} electrolyte also closely match those reported by Gullbrekken et al., with both studies using the OPLS-AA force field \cite{gullbrekken2024effect}.

\begin{table}[h]
\small
  \caption{\ RDF and coordination in ether and carbonate-based electrolytes. Cutoff for RDF is the maximum peak; for CN, it is the first minimum after the first peak in RDF (First Solvation Shell)}
  \label{tbl:RDF_CN}
  \begin{tabular*}{0.48\textwidth}{@{\extracolsep{\fill}}lllllll}
    \hline
    Electrolyte  & M & Pair & RDF & \makecell{Cutoff \\ (\text{\AA})} & CN & \makecell{Cutoff \\ (\text{\AA})} \\
    \hline
     &  & \ce{Li-O(DPE)} & 17.25 & 1.97 & 0.77 & 2.99 \\
     &  & \ce{Li-O(FSI)} & 106.15 & 2.09 & 3.82 & 3.00 \\
    DPE & 1 M & \ce{Li-Li} & 9.84 & 5.83 & 3.95 & 6.93 \\
     &  & \ce{Li-N(FSI)} & 23.25 & 4.46 & 3.47 & 5.14 \\
    & & & & & & \\
     &  & \ce{Li-O(DPE)} & 17.44 & 1.97 & 0.80 & 3.00 \\
     &  & \ce{Li-O(FSI)} & 61.99 & 2.09 & 3.81 & 3.00 \\
    DPE & 1.8 M & \ce{Li-Li} & 5.50 & 5.43 & 3.86 & 7.01 \\
     &  & \ce{Li-N(FSI)} & 13.48 & 4.50 & 3.43 & 5.26 \\
    & & & & & & \\
     &  & \ce{Li-O(DPE)} & 20.64 & 1.97 & 0.77 & 3.00 \\
     &  & \ce{Li-O(FSI)} & 33.45 & 2.08 & 3.75 & 3.00 \\
    DPE & 4 M & \ce{Li-Li} & 2.56 & 6.00 & 3.94 & 7.05 \\
     &  & \ce{Li-N(FSI)} & 7.30 & 4.45 & 3.33 & 5.23 \\
    & & & & & & \\
    &  & \ce{Li-O(FEME)} & 7.70 & 2.02 & 0.72 & 2.98 \\
     &  & \ce{Li-O(FSI)} & 103.43 & 2.09 & 3.92 & 2.97 \\
    FEME & 1 M & \ce{Li-Li} & 9.57 & 5.23 & 5.34 & 7.27 \\
     &  & \ce{Li-N(FSI)} & 28.48 & 4.49 & 3.90 & 5.56 \\
    & & & & & & \\
    &  & \ce{Li-O(FEME)} & 7.98 & 2.02 & 0.74 & 3.00 \\
     &  & \ce{Li-O(FSI)} & 62.14 & 2.09 & 3.96 & 3.00 \\
    FEME & 1.8 M & \ce{Li-Li} & 5.76 & 5.94 & 5.03 & 7.11 \\
     &  & \ce{Li-N(FSI)} & 17.15 & 4.49 & 3.86 & 5.31 \\
    & & & & & & \\
    &  & \ce{Li-O(FEME)} & 9.69 & 2.02 & 0.79 & 3.00 \\
     &  & \ce{Li-O(FSI)} & 31.84 & 2.09 & 3.91 & 3.00 \\
    FEME & 4 M & \ce{Li-Li} & 2.85 & 5.93 & 4.96 & 7.22 \\
     &  & \ce{Li-N(FSI)} & 7.93 & 4.49 & 3.63 & 5.19 \\
    & & & & & & \\
     &  & \ce{Li-O(EC)} & 8.41 & 2.11 & 0.90 & 3.49 \\
     &  & \ce{Li-O(DEC)} & 140.33 & 1.99 & 3.58 & 2.99 \\
    1:1 EC/DEC & 1 M & \ce{Li-F(PF6)} & 6.86 & 2.06 & 0.56 & 2.91 \\
    &  & \ce{Li-Li} & 0.26 & 5.93 & 0.09 & 7.32 \\
     &  & \ce{Li-P(PF6)} & 6.54 & 3.57 & 0.38 & 4.79 \\
    \hline
  \end{tabular*}
\end{table}

\begin{figure*}[ht!]
\centering
\begin{minipage}{.33\textwidth}
  \centering
  \subcaption*{a}
  \includegraphics[width=\linewidth]{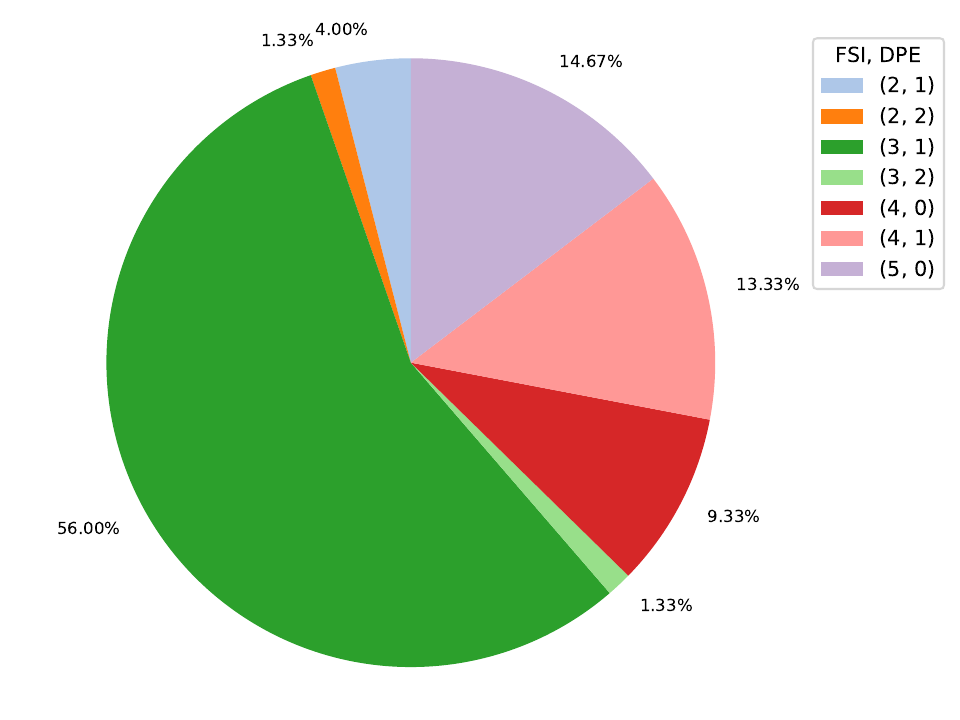}
  \phantomcaption
  \label{fig:DPE_1M_percentage_electrolyte}
\end{minipage}%
\hfill
\begin{minipage}{.33\textwidth}
  \centering
  \subcaption*{b}
  \includegraphics[width=\linewidth]{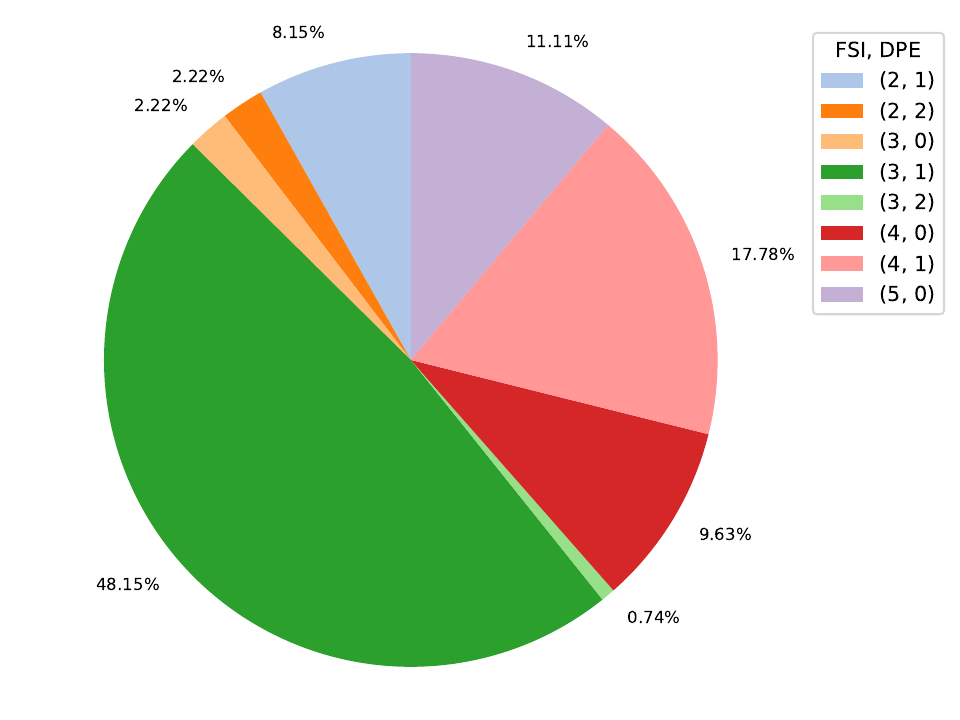}
  \phantomcaption
  \label{fig:DPE_1.8M_percentage_electrolyte}
\end{minipage}%
\hfill
\begin{minipage}{.33\textwidth}
  \centering
  \subcaption*{c}
  \includegraphics[width=\linewidth]{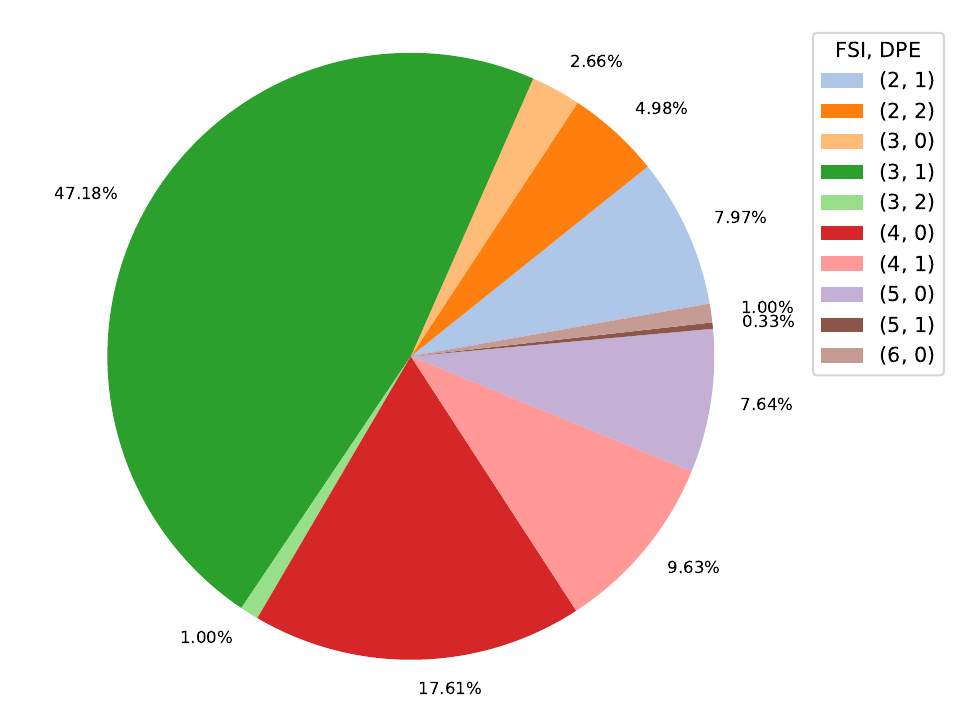}
  \phantomcaption
  \label{fig:DPE_4M_percentage_electrolyte}
\end{minipage}%
\hfill
\begin{minipage}{.33\textwidth}
  \centering
  \subcaption*{d}
  \includegraphics[width=\linewidth]{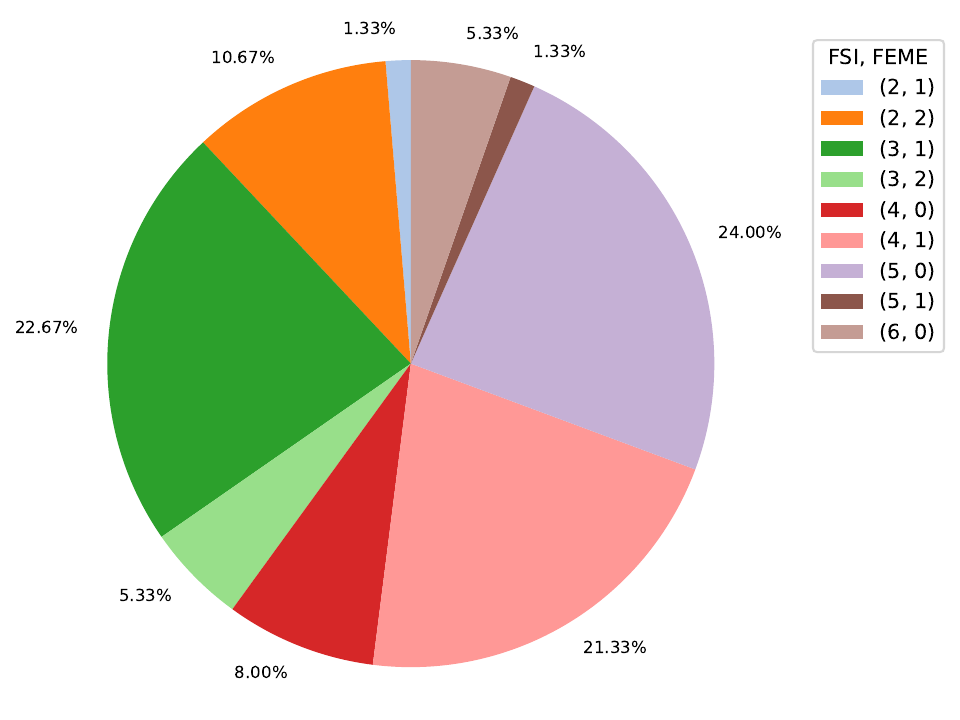}
  \phantomcaption
  \label{fig:FEME_1M_percentage_electrolyte}
\end{minipage}%
\hfill
\begin{minipage}{.33\textwidth}
  \centering
  \subcaption*{e}
  \includegraphics[width=\linewidth]{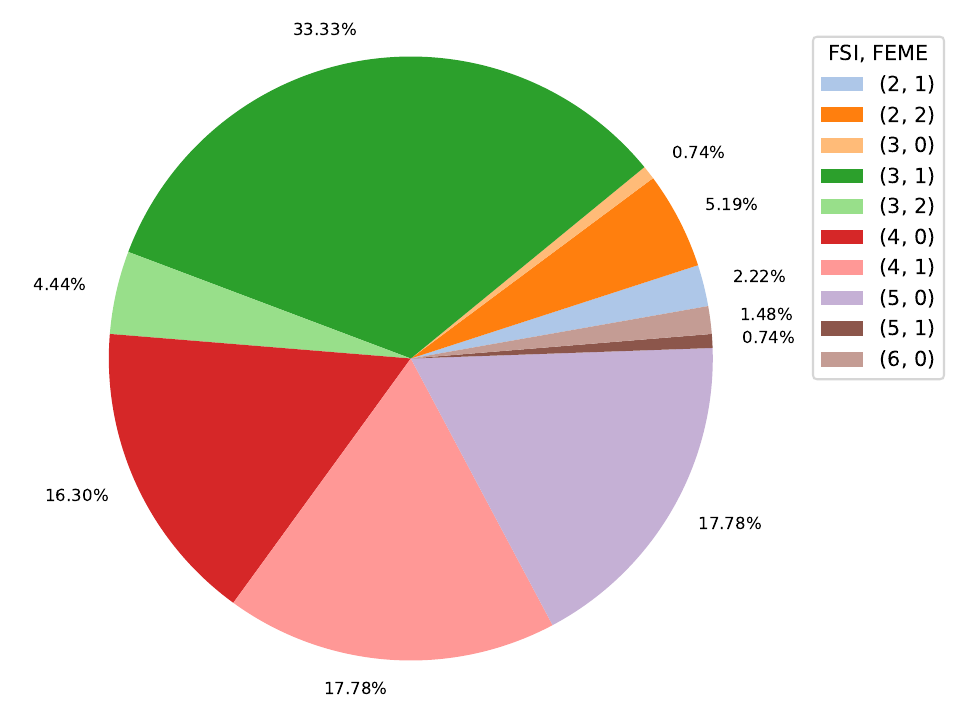}
  \phantomcaption
  \label{fig:FEME_1.8M_percentage_electrolyte}
\end{minipage}%
\hfill
\begin{minipage}{.33\textwidth}
  \centering
  \subcaption*{f}
  \includegraphics[width=\linewidth]{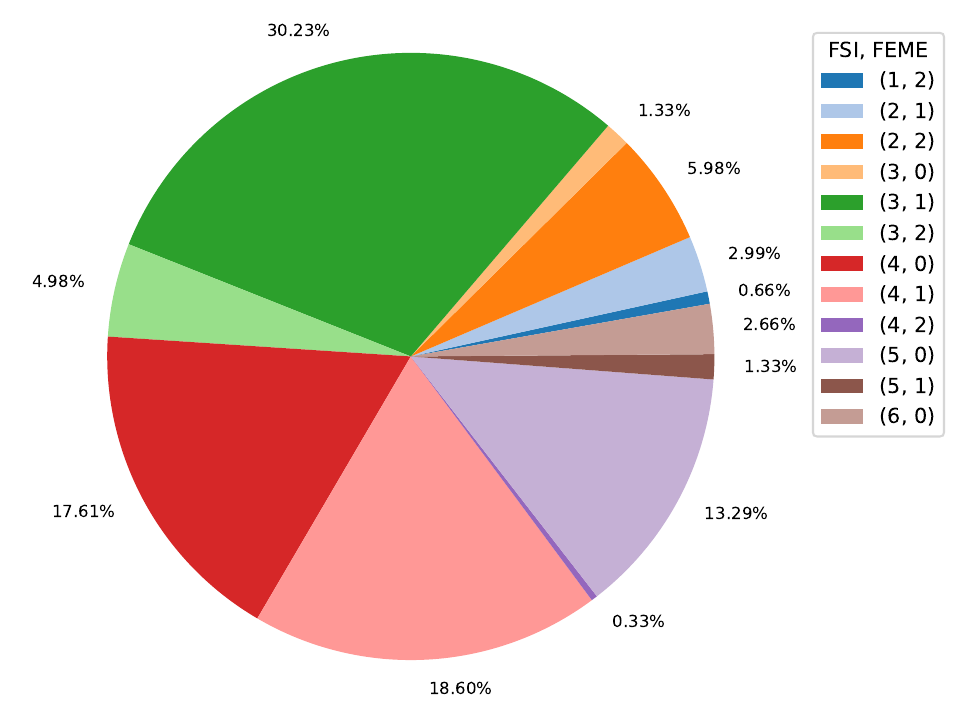}
   \phantomcaption
  \label{fig:FEME_4M_percentage_electrolyte}
\end{minipage}
\caption{Percentage of unique \ce{Li+} solvation structures in each fluorinated electrolyte: (a) DPE+1 M \ce{LiFSI}, (b) DPE+1.8 M \ce{LiFSI}, (c) DPE+4 M \ce{LiFSI}, (d) FEME+1 M \ce{LiFSI}, (e) FEME+1.8 M \ce{LiFSI}, and (f) FEME+4 M \ce{LiFSI}. The \ce{Li+} is surrounded by \ce{FSI-}, DPE and \ce{FSI-}, FEME.}
\label{fig:piechart_DPEFEME}
\end{figure*}

\begin{figure}[ht!]
\hspace{-0.15\textwidth} 
\centering
\begin{minipage}[t]{.33\textwidth}
  \centering
  \subcaption*{a}  
  \includegraphics[width=\linewidth]{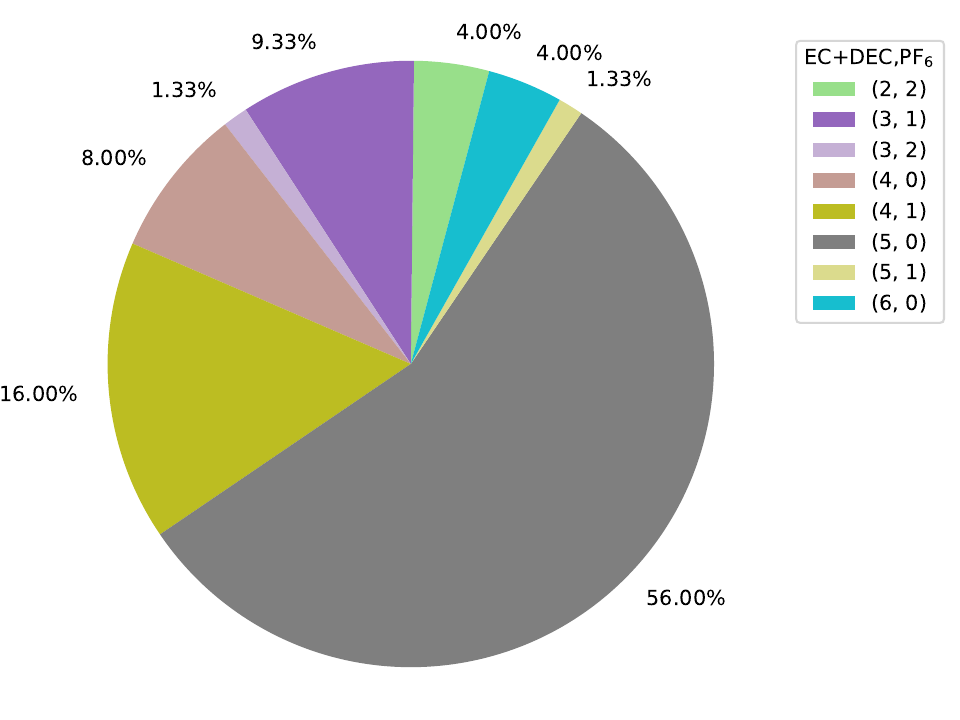}
   \phantomcaption
  \label{fig:1EC1DEC_1M_percentage_electrolyte}
  \begin{picture}(0,0)
    \put(85, 90){ 
      \begin{minipage}[t]{0.45\linewidth} 
        \centering
        \subcaption*{b} 
        \includegraphics[width=\linewidth]{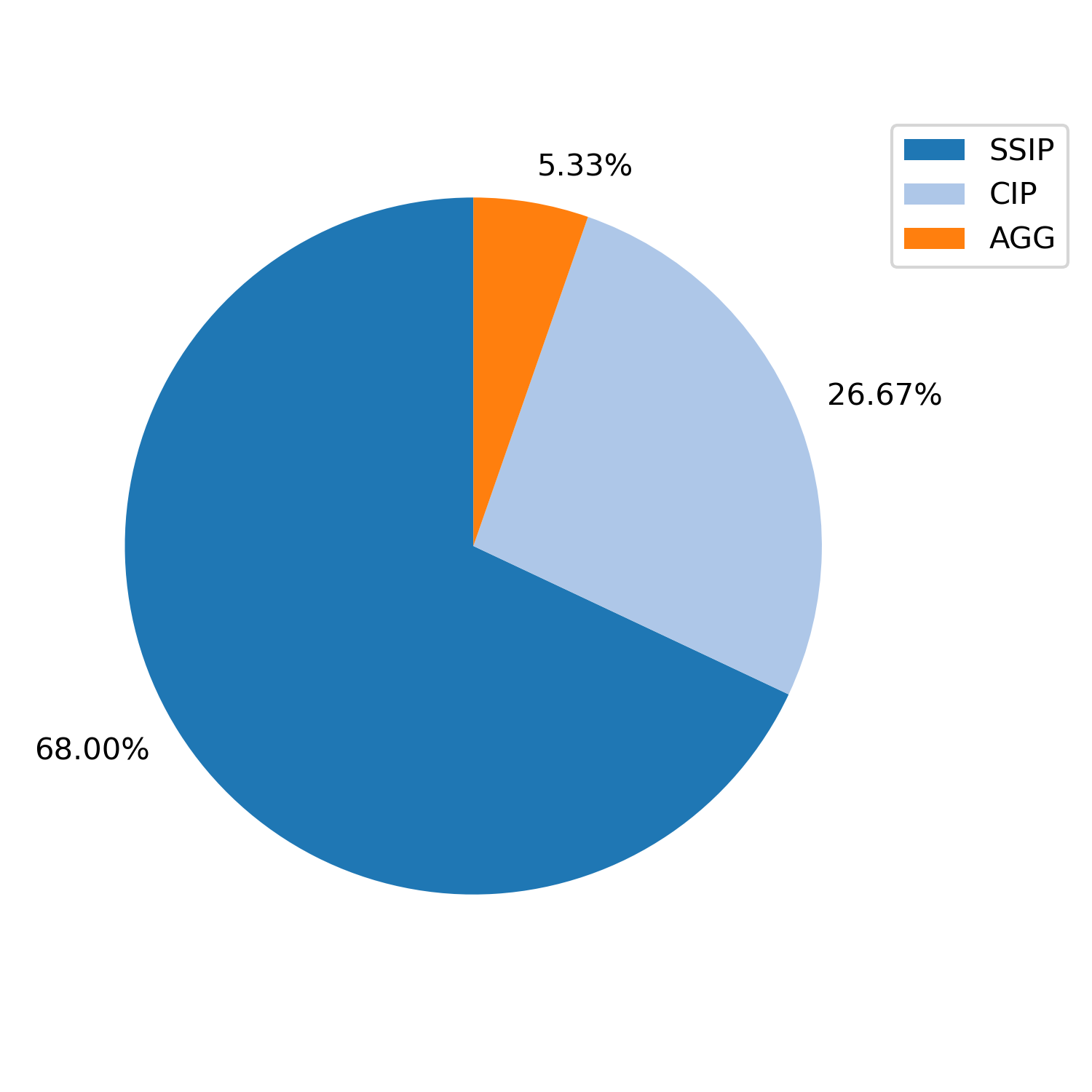}
        \phantomcaption
        \label{fig:1EC1DEC_1M_SSIPCIPAGG} 
      \end{minipage}
    }
  \end{picture}
\end{minipage}%
\caption{(a) Percentage of unique \ce{Li+} solvation structures in mixed carbonate 1:1 EC/DEC+1 M \ce{LiPF6} electrolyte. The \ce{Li+} is surrounded by EC, DEC, \ce{PF6-}. A value of \ce{PF6} = 0 indicates that the \ce{PF6-} anion is located outside the primary solvation shell (SSIPs). Inside the primary solvation shell, \ce{PF6} = 1 refers to CIPs, and \ce{PF6} $>$ 1 refers to AGGs. (b) Percentage of SSIPs, CIPs, and AGGs.}
\label{fig:piechart_ECDEC}
\end{figure}

\subsection{Solvation Structure}

\begin{table}[h]
\small
  \caption{\ Top two most dominant solvation structures and their occurrence rates, calculated from MD simulations, in fluorinated and carbonate-based electrolytes with \ce{LiFSI} and \ce{LiPF6} salt at 25°C}
  \label{tbl:percentage}
  \begin{tabular*}{0.48\textwidth}{@{\extracolsep{\fill}}lll}
    \hline
        Electrolyte System & Solvation Structure & Percentage (\%) \\
        \hline
        DPE+1 M \ce{LiFSI} & \ce{Li+\ce{(FSI-})3(DPE)1} & 56.00 \\
                              & \ce{Li+\ce{(FSI-})5(DPE)0} & 14.67 \\
        DPE+1.8 M \ce{LiFSI} & \ce{Li+\ce{(FSI-})3(DPE)1} & 48.15  \\
                                & \ce{Li+\ce{(FSI-})4(DPE)1} & 17.78 \\
        DPE+4 M \ce{LiFSI} & \ce{Li+\ce{(FSI-})3(DPE)1} & 47.18 \\
                              & \ce{Li+\ce{(FSI-})4(DPE)0} & 17.61 \\
        FEME+1 M \ce{LiFSI} & \ce{Li+\ce{(FSI-})5(FEME)0} & 24.00 \\
                                & \ce{Li+\ce{(FSI-})3(FEME)1} & 22.67  \\
        FEME+1.8 M \ce{LiFSI} & \ce{Li+\ce{(FSI-})3(FEME)1} & 33.33 \\
                     & \ce{Li+\ce{(FSI-})4(FEME)1} & 17.78 \\
        FEME+4 M \ce{LiFSI} & \ce{Li+\ce{(FSI-})3(FEME)1} & 30.23 \\
                      & \ce{Li+\ce{(FSI-})4(FEME)1}  & 18.60 \\
        1:1 EC/DEC+1 M \ce{LiPF6} & \ce{Li+\ce{(PF6-})0(EC/DEC)5} & 56.00 \\
                              & \ce{Li+\ce{(PF6-})1(EC/DEC)4} & 16.00 \\
        \hline
  \end{tabular*}
\end{table}

\begin{table}[h]
\small
\caption{\ Solvation structures and their occurrence rates, calculated from MD simulations. A value of \ce{PF6} = 0 indicates that the \ce{PF6-} anion is located outside the primary solvation shell (SSIPs). Inside the primary solvation shell, \ce{PF6} = 1 refers to CIPs, and \ce{PF6} $>$ 1 refers to AGGs}
\label{tbl:ECDEC_counting}
\begin{tabular*}{0.48\textwidth}{@{\extracolsep{\fill}}lll}
\hline
\makecell{Solvation Structures \\ EC, DEC, \ce{PF6}} & Frequency of Occurrence & Percentage (\%) \\ 
\hline
(1, 4, 0) & 25 & 33.33 \\ 
(0, 5, 0) & 10 & 13.33 \\ 
(1, 3, 1) & 7 & 9.33 \\ 
(2, 3, 0) & 7 & 9.33 \\ 
(0, 3, 1) & 5 & 6.67 \\ 
(0, 4, 1) & 4 & 5.33 \\ 
(0, 4, 0) & 3 & 4.00 \\ 
(1, 3, 0) & 3 & 4.00 \\ 
(0, 2, 2) & 3 & 4.00 \\ 
(1, 2, 1) & 2 & 2.67 \\ 
(4, 2, 0) & 2 & 2.67 \\ 
(3, 3, 0) & 1 & 1.33 \\ 
(3, 2, 1) & 1 & 1.33 \\ 
(2, 2, 1) & 1 & 1.33 \\ 
(0, 3, 2) & 1 & 1.33 \\ 
\hline
\end{tabular*}
\end{table}

\begin{figure}[h]
 \centering
 \includegraphics[width=.48\textwidth]{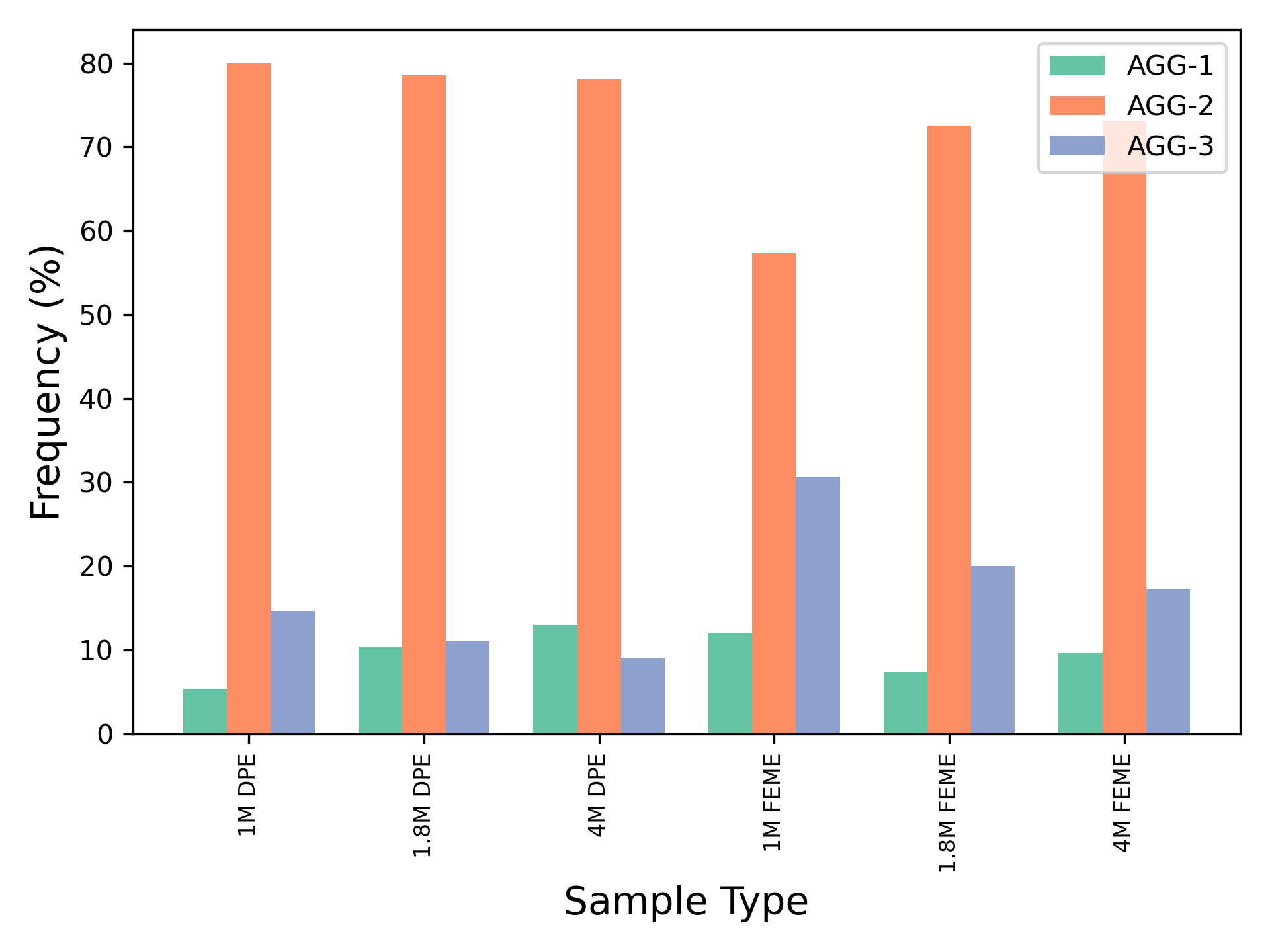}
 \caption{Detailed speciation of AGG in each fluorinated electrolyte with the frequency of occurrence. AGG-1 (one and two \ce{FSI-} anions), AGG-2 (three and four \ce{FSI-} anions), and AGG-3 (five and six \ce{FSI-} anions).}
 \label{fig:bar_AGG}
\end{figure}

\begin{figure}[h]
 \centering
 \includegraphics[width=.48\textwidth]{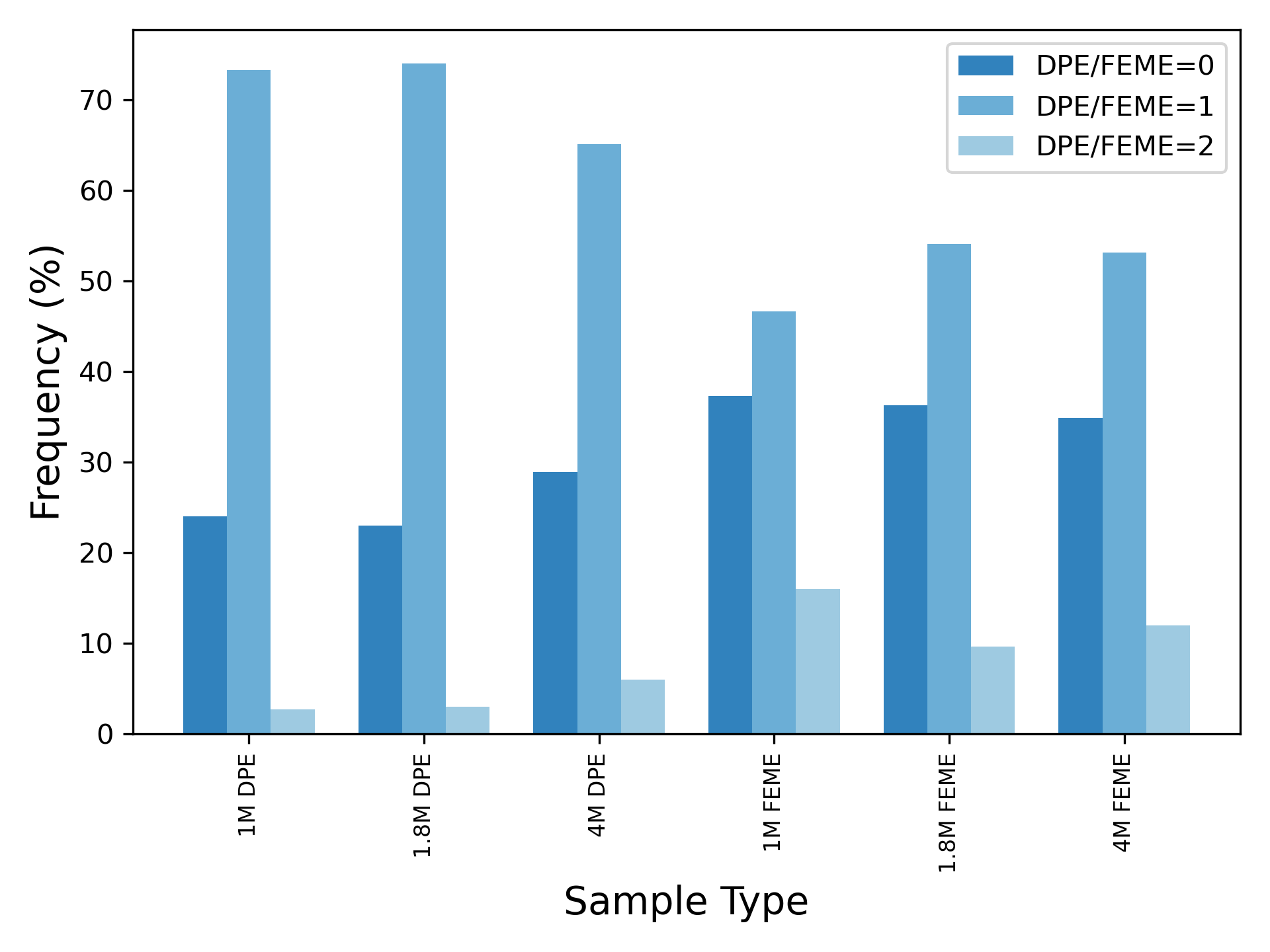}
 \caption{Frequency of occurrence of the \ce{Li+} solvation structures in each fluorinated electrolyte based on the presence of zero, one, and two solvent molecules.}
 \label{fig:bar_SOLVENT}
\end{figure}

In this work, all lithium-ion solvation structures were obtained from MD snapshots. These solvation structures significantly influence ion transport, stability, and chemical reactivity in electrolytes, which are critical for the design of lithium-ion batteries. The solvation structures in DPE- and FEME-based electrolytes are categorized into three groups: AGG-1 (containing one or two \ce{FSI-} anions), AGG-2 (three or four \ce{FSI-} anions), and AGG-3 (five or six \ce{FSI-} anions). In the mixed EC/DEC electrolyte, the solvation structures are classified as solvent-separated ion pairs (SSIPs), contact ion pairs (CIPs), and aggregates (AGGs). The solvation structures in DPE+1.8 M \ce{LiFSI}, FEME+1.8 M \ce{LiFSI}, and EC/DEC+1 M \ce{LiPF6} electrolytes are illustrated in Fig. S15\dag\ to S19\dag. The frequency of occurrence of all possible unique solvation structures in each electrolyte is analyzed in Fig. \ref{fig:piechart_DPEFEME} and \ref{fig:piechart_ECDEC}. Fig. \ref{fig:1EC1DEC_1M_SSIPCIPAGG} also shows the percentage of SSIP, CIP, and AGG in mixed EC/DEC electrolyte. The two most frequent solvation structures in each electrolyte and their respective percentages are shown in Table \ref{tbl:percentage}. In DPE+1.8 M \ce{LiFSI} electrolyte, the two most dominant solvation structures are \ce{Li+\ce{(FSI-})3(DPE)1} (48.15\%) and \ce{Li+\ce{(FSI-})4(DPE)1} (17.78\%). Similarly, in FEME+1.8 M \ce{LiFSI} electrolyte, the two most dominant solvation structures are \ce{Li+\ce{(FSI-})3(FEME)1} (33.33\%) and \ce{Li+\ce{(FSI-})4(FEME)1} (17.78\%). According to our study, the primary solvation structure remains nearly unchanged across varying salt concentrations in each DPE- and FEME-based electrolyte. Table \ref{tbl:ECDEC_counting} provides a thorough analysis of each solvation structure in the mixed EC/DEC electrolyte, specifying the individual counts of EC and DEC solvent molecules. Our findings also indicate that the number of unique solvation structures in DPE+\ce{LiFSI} and FEME+\ce{LiFSI} electrolytes increases as salt concentration increases \cite{yang2023first}. In DPE+\ce{LiFSI}, as the salt concentration rises from 1 M to 1.8 M to 4 M, the number of unique solvation structures increases from 7 to 8 to 10 (Fig. \ref{fig:DPE_1M_percentage_electrolyte}, \ref{fig:DPE_1.8M_percentage_electrolyte}, and \ref{fig:DPE_4M_percentage_electrolyte}). Similarly, in FEME+\ce{LiFSI}, increasing the concentration from 1 M to 1.8 M to 4 M results in an increase in unique solvation structures from 9 to 10 to 12 (Fig. \ref{fig:FEME_1M_percentage_electrolyte}, \ref{fig:FEME_1.8M_percentage_electrolyte}, and \ref{fig:FEME_4M_percentage_electrolyte}). However, each FEME+\ce{LiFSI} electrolyte exhibits a slightly higher number of possible solvation structures than DPE+\ce{LiFSI} electrolytes. A detailed speciation of AGG in each DPE- and FEME-based electrolyte is shown in Fig. \ref{fig:bar_AGG}. At all salt concentrations, in DPE+\ce{LiFSI} and FEME+\ce{LiFSI} electrolytes, the primary aggregates belong to the AGG-2 category, containing three or four \ce{FSI-} anions and one solvent molecule (Fig. \ref{fig:bar_AGG} and \ref{fig:bar_SOLVENT}). Our results show that in each DPE- and FEME-based electrolyte, the solvation structures are primarily AGG (Fig. \ref{fig:piechart_DPEFEME}), whereas in the mixed EC/DEC electrolyte, 68\% are SSIP and only 5.33\% are AGG (Fig. \ref{fig:1EC1DEC_1M_SSIPCIPAGG}).

\begin{figure*}[ht!]
\centering
\begin{subfigure}{.5\textwidth}
  \centering
  \includegraphics[width=\linewidth]{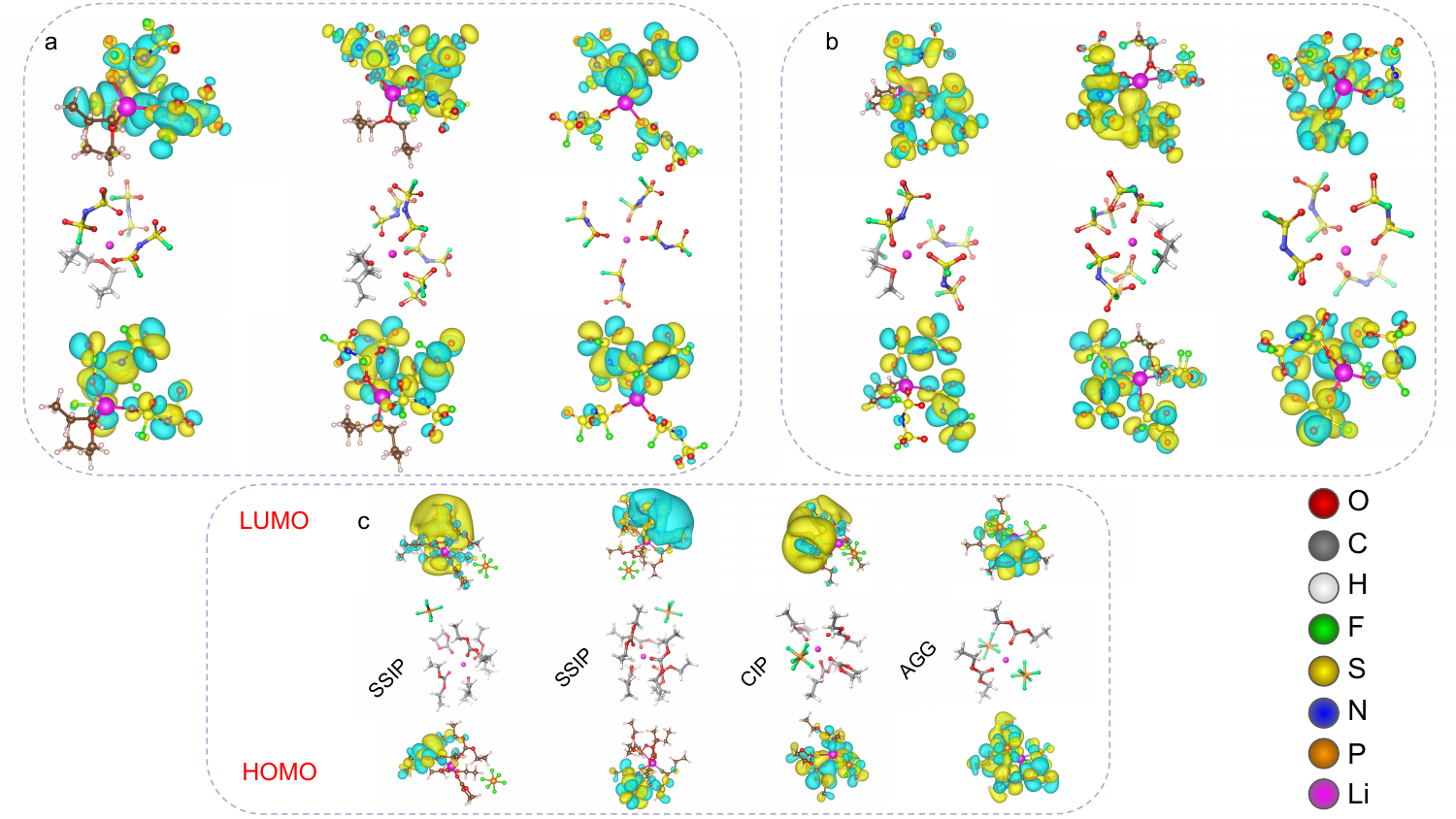}
  \label{fig:HOMO_LUMO_cluster}
\end{subfigure}%
\hfill
\begin{subfigure}{.5\textwidth}
  \centering
  \includegraphics[width=\linewidth]{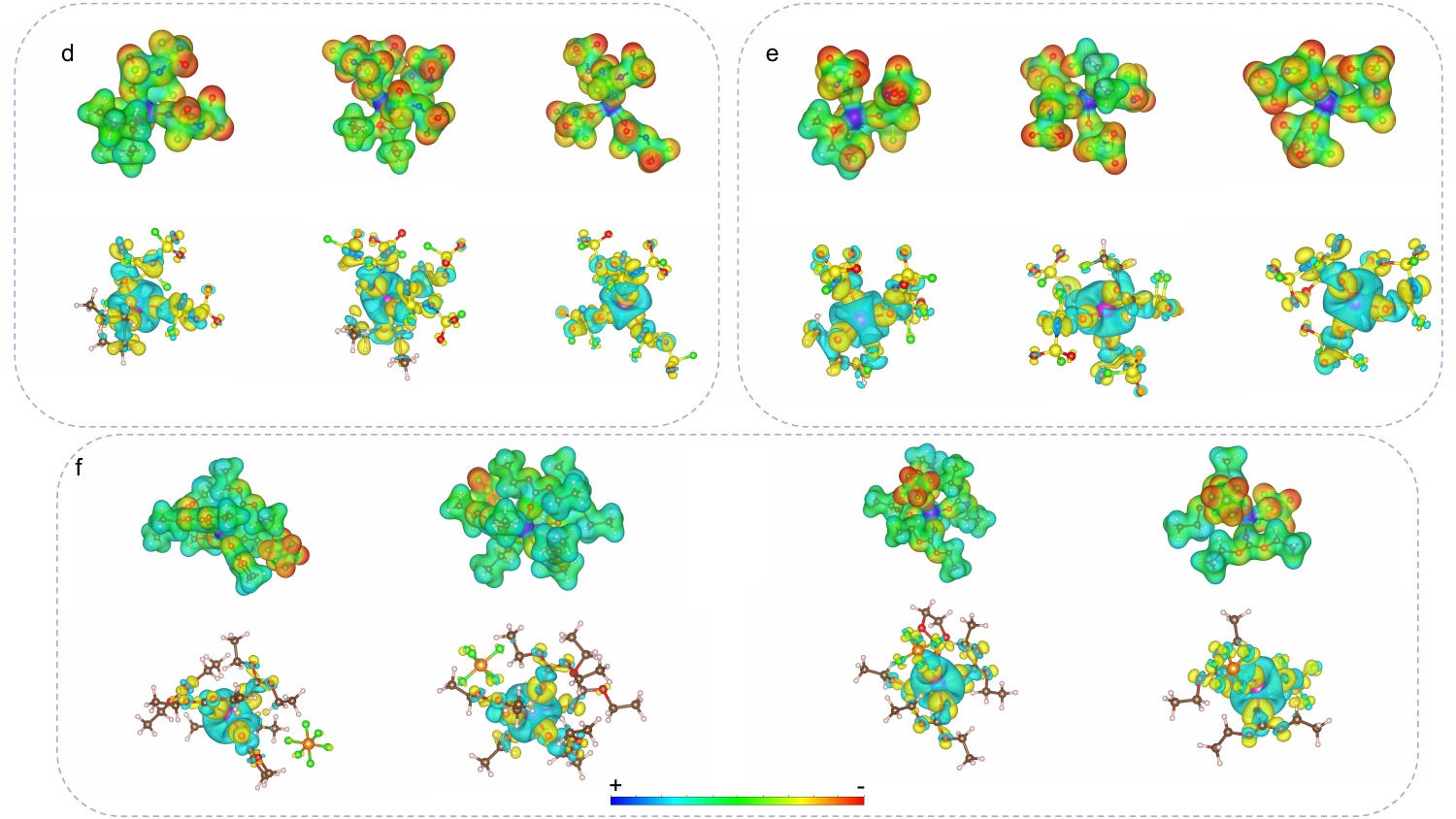}
  \label{fig:ESP_CDD_cluster}
\end{subfigure}
\caption{\textbf Top two most solvated structures (\ce{Li+\ce{(FSI-})3(DPE)1}, \ce{Li+\ce{(FSI-})4(DPE)1}, \ce{Li+\ce{(FSI-})3(FEME)1}, \ce{Li+\ce{(FSI-})4(FEME)1}, \ce{Li+\ce{(PF6-})0(EC)1(DEC)4}, and \ce{Li+\ce{(PF6-})0(EC)0(DEC)5} including \ce{Li+\ce{(FSI-})4(DPE)0}, \ce{Li+\ce{(FSI-})4(FEME)0}), CIP, and AGG with their HOMO/LUMO and corresponding ESP and CDD in (a), (d) DPE+1.8 M \ce{LiFSI}, (b), (e) FEME+1.8 M \ce{LiFSI}, and (c), (f) EC/DEC+1 M \ce{LiPF6} electrolytes. The positive and negative phase of HOMO and LUMO are depicted in yellow and cyan colors , respectively. Yellow and cyan indicate the different signs of the isosurface of the wave function, and their sizes indicate its amplitude. In the ESP maps, the red and blue regions represent areas of high electron density (negative charge) and low electron density (positive charge), respectively. In the CDD plot, the cyan region represents electron depletion and the yellow region represents electron accumulation. The isosurfaces of HOMO/LUMO, ESP, and CDD were visualized using VESTA. The isosurface levels were set between $1 \times 10^{-10}$ and $1 \times 10^{-8}$ for HOMO/LUMO, $10$ for ESP, and $0.0009$ for CDD. HOMO/LUMO diagrams are shown in Figures (a), (b), and (c), while ESP maps and CDD plots are presented in Figures (d), (e), and (f).}
\label{fig:ESP_CDD_HOMOLUMO}
\end{figure*}

\ce{Li+} diffusion in electrolytes mainly happens through two types of mechanisms: the vehicle mechanism and the hopping mechanism \cite{zhou2023strategies, kramer2025importance}. In the vehicle mechanism, which is common in electrolytes with moderate salt concentration, \ce{Li+} ions move through the electrolyte where the \ce{Li+} ion and its coordinated solvent molecules migrate together as a complex. This can usually leads to a lower \ce{Li+} transference number because the ion’s movement is tied to the motion of the solvent and anions. In contrast, hopping mechanism is prevalent in concentrated electrolytes \cite{dokko2018direct, yu2023uncorrelated, ugata2021structural, kondou2019ionic, ugata2019li, galle2020molecular}. Here, \ce{Li+} ions move by jumping between different coordination sites in the electrolyte. These sites can be either solvent molecules or anions. This hopping mechanism often leads to higher \ce{Li+} mobility and transference numbers, as \ce{Li+} ions move more independently from the bulk solvent. Our MD and DFT studies show clear differences in \ce{Li+} diffusion mechanism depending on the solvation structure. The \ce{Li+} ions in the AGG-dominated DPE and FEME systems are likely to favor a hopping-type diffusion mechanism \cite{zhou2023strategies, kramer2025importance}. According to Saito et al., the increased microviscosity in WSEE, caused by strong cation-anion and cation-polymer interactions, slows down ion diffusion \cite{saito2017selective}. This, along with aggregate formation via bridging coordination, leads to suppressed ionic transport and reduced conductivity in the DPE electrolyte as confirmed by Li et al. \cite{li2023non} Conversely, in the SSIP-dominated mixed EC/DEC system, the \ce{Li+} ions are strongly solvated by solvent molecules and likely diffuse via a vehicle mechanism \cite{zhou2023strategies, kramer2025importance}. These differences in solvation structure and diffusion mechanism directly influence ionic conductivity, \ce{Li+} transference numbers, and ultimately battery performance. In future work, more electrolyte properties such as ionic conductivity, voltage window, and \ce{Li+} transference number can be explored \cite{li2023non, lorenz2024evaluating}. These properties will further validate the transport behavior suggested by the observed solvation structures.

The charge density difference, Bader charge analysis, electrostatic potential maps, and binding energies of the solvation structures are discussed in the following sections.

\subsubsection{Charge Density Difference and Bader Charge Analysis with ESP Maps.~~}

In addition to Bader atomic charge calculations in VASP, the charge density difference (CDD) was also calculated for all the \ce{Li+} solvation structures using Equation \ref{Eq:CDD} \cite{choudhuri2020calculating}. Here, $\rho_{\text{total system}}$, $\rho_{\text{subsystem1}}$, and $\rho_{\text{subsystem2}}$ represent the charge densities of the \ce{Li+} solvation structure (\ce{Li}+solvents+anions), \ce{Li}, and the combined solvents and anions, respectively. These calculations, including CDD and Bader charge analysis, were performed to compute the amount of charge transfer from the \ce{Li+} ion to the surrounding solvents and anions and to analyze their electronic distribution \cite{zhang2021blue}. Fig. \ref{fig:ESP_CDD_HOMOLUMO} shows the CDD plot of \ce{Li+} solvation structures, where the yellow regions indicate electron gain and the cyan regions represent electron loss. These CDD plots depict the regions of electron loss around the \ce{Li+} ion and electron accumulation between the ionized \ce{Li} atom and the \ce{O} atoms, indicating charge transfer from the ionized \ce{Li} atom to the \ce{O} atoms. This suggests a strong binding interaction between \ce{Li} and the \ce{O} atoms of the anions and solvents in the \ce{Li+} solvation structures. In the case of CIPs and AGGs in the EC/DEC electrolyte, strong binding interactions occur between \ce{Li} and both the \ce{F} atom of the \ce{PF6-} anion and the \ce{O} atom of the solvents. The amount of charge transfer from the \ce{Li+} ion to the surrounding solvents and anions was calculated using Bader charge analysis and is listed in Tables \ref{tbl:Li_solvent/anion}, \ref{tbl:solvatin_structuer_DPE}, \ref{tbl:solvatin_structuer_FEME}, \ref{tbl:solvatin_structuer_ECDEC}, and \ref{tbl:solvatin_structuer_ECDEC_noPF6}. Reduced charge transfer is observed in \ce{Li+}-carbonate systems compared to \ce{Li+}-ether systems (Table \ref{tbl:Li_solvent/anion}). Approximately $0.70$ - $0.86$ $e$ (Table \ref{tbl:solvatin_structuer_DPE}) and $0.67$ - $0.79$ $e$ (Table \ref{tbl:solvatin_structuer_FEME}) are transferred from \ce{Li+} to the surrounding solvents and anions in the AGGs of the 1.8 M DPE and 1.8 M FEME electrolytes. Similarly, approximately $0.79$ - $0.85$ $e$, $0.71$ - $0.84$ $e$, and $0.80$ $e$ (Table \ref{tbl:solvatin_structuer_ECDEC}) are transferred from \ce{Li+} to the surrounding solvents and anions in the SSIPs, CIPs, and AGGs of the 1 M EC/DEC electrolyte. Additionally, the ESP maps of these \ce{Li+} solvation structures in Fig. \ref{fig:ESP_CDD_HOMOLUMO} show that the negative charge is primarily localized on the \ce{O} and \ce{F} atoms, while the positive charge is mainly localized on the \ce{Li} atom.

\begin{gather}
  \label{Eq:CDD}
  \Delta \rho = \rho_{\text{total system}} - \rho_{\text{subsystem1}} -\rho_{\text{subsystem2}}
\end{gather}

\begin{table*}[ht!]
\small
  \caption{\ Calculated quantities of \ce{Li+-solvent} and \ce{Li+-anion} systems}
  \label{tbl:Li_solvent/anion}
  \begin{tabular*}{\textwidth}{@{\extracolsep{\fill}}lllllll}
    \hline
    Systems  & HOMO (eV) & LUMO (eV) & LUMO–HOMO (eV) & Binding Energy (eV) & Bader Charge of \ce{Li}, $q$ ($e$) & $\Delta q (e)$ \\
    \hline
    \ce{Li+(DPE)1} & -1.687 & -0.638 & 1.05 & -0.561 & +0.726 & +0.274 \\
    \ce{Li+(FEME)1} & -2.002 & -0.806 & 1.20 & -0.531 & +0.730 & +0.270 \\
    \ce{Li+\ce{(FSI-})1} & -6.270 & -0.993 & 5.28 & -5.431 & +0.160 & +0.840 \\
    \ce{Li+(EC)1} & -1.742 & -1.394 & 0.35 & -0.651 & +0.788 & +0.212 \\
    \ce{Li+(DEC)1} & -1.698 & -1.156 & 0.54 & -0.671 & +0.801 & +0.199 \\
    \ce{Li+\ce{(PF6-})1} & -8.077 & -1.158 & 6.92 & -7.491 & +0.053 & +0.947 \\
    \hline
  \end{tabular*}
\end{table*}

\begin{table*}[ht!]
\small
  \caption{\ Calculated quantities of \ce{Li+} solvation structures in DPE+1.8 M \ce{LiFSI} electrolyte}
  \label{tbl:solvatin_structuer_DPE}
  \begin{tabular*}{\textwidth}{@{\extracolsep{\fill}}llllllll}
    \hline
    Solvation Structure  & Species & HOMO (eV) & LUMO (eV) & LUMO–HOMO (eV) & Binding Energy (eV) & Bader Charge of \ce{Li}, $q$ ($e$) & $\Delta q (e)$ \\
    \hline
    \ce{Li+\ce{(FSI-})2(DPE)1} & AGG1 & -6.949 & -0.940 & 6.01 & -7.901 & +0.146 & +0.854 \\
    \ce{Li+\ce{(FSI-})2(DPE)2} & AGG1 & -6.076 & -0.482 & 5.59 & -8.331 & +0.141 & +0.859 \\
    \ce{Li+\ce{(FSI-})3(DPE)0} & AGG2 & -7.279 & -1.554 & 5.73 & -7.291 & +0.202 & +0.798 \\
    \ce{Li+\ce{(FSI-})4(DPE)0} & AGG2 & -6.981 & -1.350 & 5.63 & -7.501 & +0.195 & +0.805 \\
    \ce{Li+\ce{(FSI-})3(DPE)1} & AGG2 & -6.590 & -0.966 & 5.62 & -8.011 & +0.299 & +0.701 \\
    \ce{Li+\ce{(FSI-})4(DPE)1} & AGG2 & -6.328 & -0.838 & 5.49 & -8.601 & +0.235 & +0.765 \\
    \ce{Li+\ce{(FSI-})3(DPE)2} & AGG2 & -6.118 & -0.406 & 5.71 & -8.871 & +0.286 & +0.714 \\
    \ce{Li+\ce{(FSI-})5(DPE)0} & AGG3 & -6.364 & -2.702 & 3.66 & -8.911 & +0.251 & +0.749\\
    \hline
  \end{tabular*}
\end{table*}

\begin{table*}[ht!]
\small
  \caption{\ Calculated quantities of \ce{Li+} solvation structures in FEME+1.8 M \ce{LiFSI} electrolyte}
  \label{tbl:solvatin_structuer_FEME}
  \begin{tabular*}{\textwidth}{@{\extracolsep{\fill}}llllllll}
    \hline
    Solvation Structure  & Species & HOMO (eV) & LUMO (eV) & LUMO–HOMO (eV) & Binding Energy (eV) & Bader Charge of \ce{Li}, $q$ ($e$) & $\Delta q (e)$ \\
    \hline
    \ce{Li+\ce{(FSI-})2(FEME)1} & AGG1 & -6.927 & -1.101 & 5.83 & -7.621 & +0.239 & +0.761 \\
    \ce{Li+\ce{(FSI-})2(FEME)2} & AGG1 & -6.405 & -0.875 & 5.53 & -8.001 & +0.232 & +0.768 \\
    \ce{Li+\ce{(FSI-})3(FEME)0} & AGG2 & -7.200 & -1.543 & 5.66 & -7.361 & +0.226 & +0.774 \\
    \ce{Li+\ce{(FSI-})4(FEME)0} & AGG2 & -6.841 & -1.365 & 5.48 & -7.771 & +0.309 & +0.691 \\
    \ce{Li+\ce{(FSI-})3(FEME)1} & AGG2 & -6.787 & -0.972 & 5.81 & -7.761 & +0.314 & +0.686 \\
    \ce{Li+\ce{(FSI-})4(FEME)1} & AGG2 & -6.570 & -0.915 & 5.66 & -8.171 & +0.252 & +0.748 \\
    \ce{Li+\ce{(FSI-})3(FEME)2} & AGG2 & -6.448 & -0.628 & 5.82 & -8.381 & +0.227 & +0.773 \\
    \ce{Li+\ce{(FSI-})5(FEME)0} & AGG3 & -6.664 & -1.037 & 5.63 & -7.951 & +0.266 & +0.734 \\
    \ce{Li+\ce{(FSI-})6(FEME)0} & AGG3 & -6.406 & -0.829 & 5.58 & -8.261 & +0.215 & +0.785 \\
    \ce{Li+\ce{(FSI-})5(FEME)1} & AGG3 & -6.347 & -0.719 & 5.63 & -8.641 & +0.330 & +0.670 \\
    \hline
  \end{tabular*}
\end{table*}

\begin{table*}[ht!]
\small
  \caption{\ Calculated quantities of \ce{Li+} solvation structures in EC/DEC+1 M \ce{LiPF6} electrolyte}
  \label{tbl:solvatin_structuer_ECDEC}
  \begin{tabular*}{\textwidth}{@{\extracolsep{\fill}}llllllll}
    \hline
    Solvation Structure  & Species & HOMO (eV) & LUMO (eV) & LUMO–HOMO (eV) & Binding Energy (eV) & Bader Charge of \ce{Li}, $q$ ($e$) & $\Delta q (e)$ \\
    \hline
    \ce{Li+(EC)0(DEC)4} & SSIP & -6.433 & -0.860 & 5.57 & -10.151 & +0.214 & +0.786 \\
    \ce{Li+(EC)0(DEC)5} & SSIP & -6.241 & -0.808 & 5.43 & -10.591 & +0.158 & +0.842 \\
    \ce{Li+(EC)1(DEC)3} & SSIP & -6.021 & -1.216 & 4.81 & -9.4810  & +0.192 & +0.808 \\
    \ce{Li+(EC)1(DEC)4} & SSIP & -6.484 & -0.667 & 5.82 & -10.761 & +0.192 & +0.808 \\
    \ce{Li+(EC)2(DEC)3} & SSIP & -6.559 & -0.786 & 5.77 & -10.621 & +0.203 & +0.797 \\
    \ce{Li+(EC)3(DEC)3} & SSIP & -6.118 & -0.856 & 5.26 & -10.741 & +0.149 & +0.851 \\
    \ce{Li+(EC)4(DEC)2} & SSIP & -5.704 & -1.076 & 4.63 & -10.121 & +0.164 & +0.836 \\
    \ce{Li+\ce{(PF6-})1(EC)0(DEC)3} & CIP & -6.988 & -0.877 & 6.11 & -9.8710  & +0.165 & +0.835 \\
    \ce{Li+\ce{(PF6-})1(EC)0(DEC)4} & CIP & -6.061 & -0.736 & 5.33 & -10.121 & +0.289 & +0.711 \\
    \ce{Li+\ce{(PF6-})1(EC)1(DEC)2} & CIP & -6.765 & -1.003 & 5.76 & -9.7510  & +0.235 & +0.765 \\
    \ce{Li+\ce{(PF6-})1(EC)1(DEC)3} & CIP & -6.648 & -0.759 & 5.89 & -10.201 & +0.160 & +0.840 \\
    \ce{Li+\ce{(PF6-})1(EC)2(DEC)2} & CIP & -6.459 & -0.743 & 5.72 & -10.201 & +0.231 & +0.769 \\
    \ce{Li+\ce{(PF6-})1(EC)3(DEC)2} & CIP & -6.517 & -0.719 & 5.80 & -11.051 & +0.177 & +0.823 \\
    \ce{Li+\ce{(PF6-})2(EC)0(DEC)2} & AGG & -7.808 & -1.465 & 6.34 & -10.971 & +0.198 & +0.802 \\
    \ce{Li+\ce{(PF6-})2(EC)0(DEC)3} & AGG & -7.390 & -1.104 & 6.29 & -11.551 & +0.196 & +0.804 \\
    \hline
  \end{tabular*}
\end{table*}

\begin{table*}[ht!]
\small
  \caption{\ Calculated quantities of SSIPs in EC/DEC+1 M \ce{LiPF6} electrolyte. \ce{PF6-} anion is removed from SSIPs}
  \label{tbl:solvatin_structuer_ECDEC_noPF6}
  \begin{tabular*}{\textwidth}{@{\extracolsep{\fill}}llllllll}
    \hline
    Solvation Structure  & Species & HOMO (eV) & LUMO (eV) & LUMO–HOMO (eV) & Binding Energy (eV) & Bader Charge of \ce{Li}, $q$ ($e$) & $\Delta q (e)$ \\
    \hline
    \ce{Li+(EC)0(DEC)4} & SSIP & -0.912 & -0.649 & 0.26 & -2.611 & +0.179 & +0.821 \\
    \ce{Li+(EC)0(DEC)5} & SSIP & -0.756 & -0.565 & 0.19 & -3.061 & +0.151 & +0.849 \\
    \ce{Li+(EC)1(DEC)3} & SSIP & -0.947 & -0.767 & 0.18 & -2.391 & +0.207 & +0.793 \\
    \ce{Li+(EC)1(DEC)4} & SSIP & -0.867 & -0.599 & 0.27 & -3.001 & +0.187 & +0.813 \\
    \ce{Li+(EC)2(DEC)3} & SSIP & -0.891 & -0.643 & 0.25 & -2.891 & +0.189 & +0.811 \\
    \ce{Li+(EC)3(DEC)3} & SSIP & -0.838 & -0.536 & 0.30 & -3.461 & +0.176 & +0.824 \\
    \ce{Li+(EC)4(DEC)2} & SSIP & -0.793 & -0.526 & 0.27 & -3.081 & +0.204 & +0.796 \\
    \hline
  \end{tabular*}
\end{table*}

\subsubsection{Binding Energy.~~}

\begin{figure}[h]
 \centering
 \includegraphics[width=.48\textwidth]{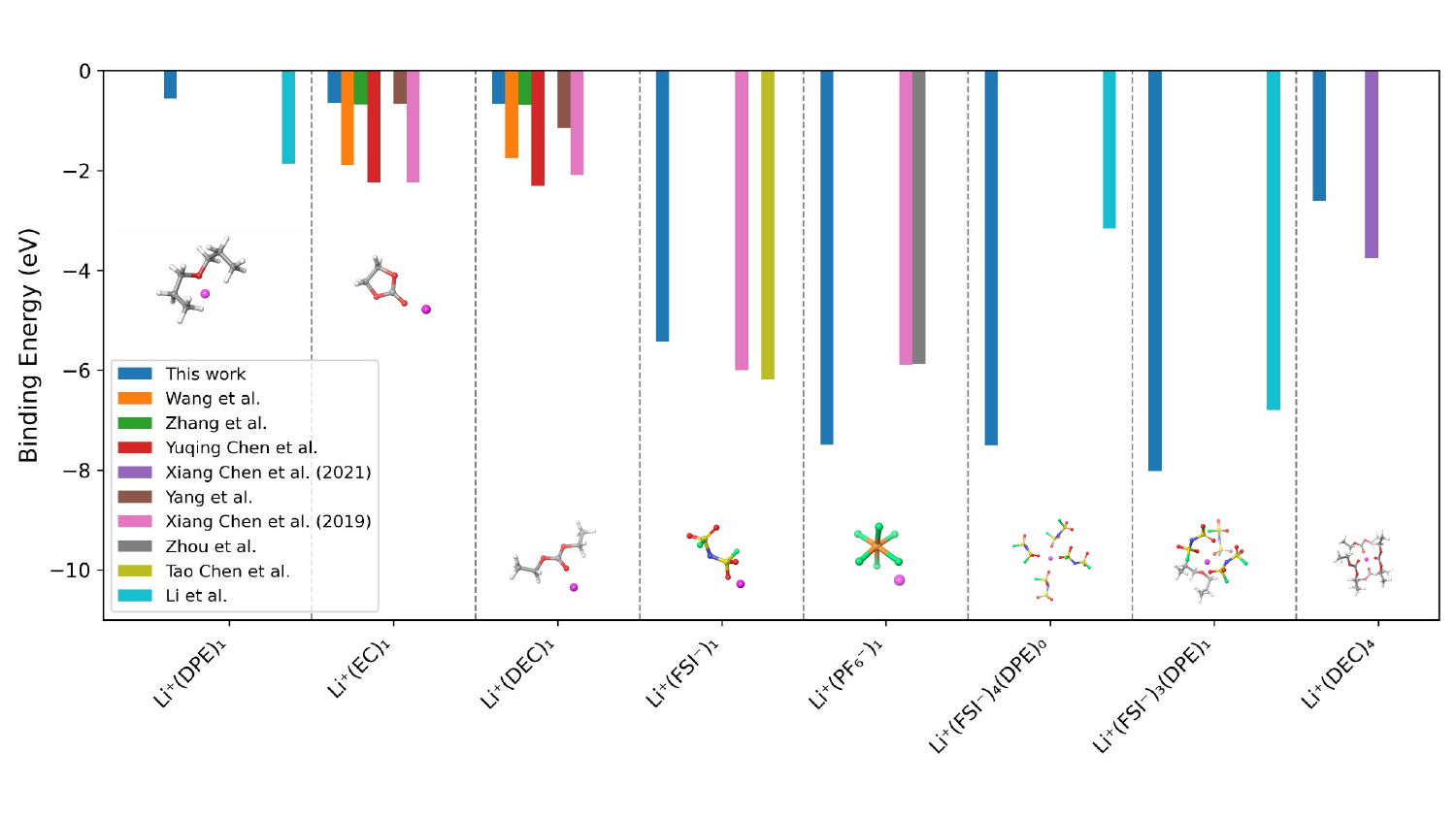}
 \caption{DFT validation of binding energies \cite{wang2021ion, zhang2023all, chen2023breaking, chen2021ion, yang2021formation, chen2019cation, zhou2025tuning, chen2022stable, li2023non}.}
 \label{fig:binding_energy_compare}
\end{figure}

The binding energies of different lithium-ion solvated systems were investigated using DFT calculations. The binding energy of the \ce{Li+} solvation structure was calculated by subtracting the energies of individual components from the total energy of the \ce{Li+} solvation structure, using Equations \ref{Eq:DPEFEME_BE} and \ref{Eq:ECDEC_BE}, where $E$ represents the energy, and $n$ and $m$ are the number of solvent and anion species in the lithium-ion solvation structure, respectively \cite{li2023non, yang2023first}. Our findings show that DPE and FEME exhibit significantly reduced binding energy to \ce{Li+} (Fig. \ref{fig:binding_energy_compare} and Table \ref{tbl:Li_solvent/anion}), which effectively suppresses salt dissociation and promotes ion aggregate formation starting at a concentration of 1 M \cite{li2023non}. The binding energies of these AGGs in the 1.8 M DPE and 1.8 M FEME electrolytes are quite similar, ranging from -8.91 to -7.29 eV and -8.64 to -7.36 eV, respectively (Tables \ref{tbl:solvatin_structuer_DPE} and \ref{tbl:solvatin_structuer_FEME}). For the SSIPs (including \ce{PF6-}), CIPs, and AGGs, the binding energy ranges from -11.55 to -9.48 eV, while for SSIPs (excluding \ce{PF6-}), the binding energy is lower, ranging from -3.46 to -2.39 eV (Tables \ref{tbl:solvatin_structuer_ECDEC} and \ref{tbl:solvatin_structuer_ECDEC_noPF6}). The DFT validation is shown in Fig. \ref{fig:binding_energy_compare}. The slight differences in binding energies for the same systems depend on the software used (VASP or Gaussian), the functional (PBE or B3LYP), the simulation box size, and the orientation of the structure \cite{wang2021ion, zhang2023all, chen2023breaking, chen2021ion, yang2021formation, chen2019cation, zhou2025tuning, chen2022stable, li2023non}.

Binding energy is the energy required to break a system into its individual components and separate them infinitely. In the context of lithium-ion solvation structures, binding energy refers to the strength of interaction between the \ce{Li+} ion and the solvent or anions species in the cation solvation shell. It quantifies how tightly the \ce{Li+} ion is bound to its surrounding environment, including solvent molecules (DPE, FEME, EC, DEC) and the counterions (\ce{FSI-}, \ce{PF6-}). If the binding energy is negative, bond formation is likely exergonic. Moreover, a higher absolute value of the binding energy reflects stronger interactions among the species \cite{atwi2022mispr}. According to the literature, during the desolvation process, \ce{Li+} ions separate from the solvated molecules, move through the solid electrolyte interphase (SEI) layer, and diffuse into the electrode \cite{wang2022structural}. The binding energy in the \ce{Li+} solvation structure is important because it affects how easily the \ce{Li+} ion moves through the electrolyte and intercalates into the electrode, which influences battery performance. Recent studies have shown that using isobutyronitrile (iBN) as a cosolvent weakens the \ce{Li+-solvent} interaction, making \ce{Li+} desolvation easier and thereby improving low-temperature ionic mobility \cite{luo2024enabling}. A higher binding energy (greater stability of solvation) means stronger interactions between the \ce{Li+} ion and surrounding molecules, making it harder to desolvate the \ce{Li+} ion and slowing battery performance, especially at low temperatures \cite{fu2018correlating}. Conversely, lower binding energy allows the \ce{Li+} ion to move more easily, improving battery efficiency. From Fig. \ref{fig:binding_energy_compare} and Table \ref{tbl:Li_solvent/anion}, the binding energy values indicate that DEC (-0.671 eV) and EC (-0.651 eV) exhibit stronger solvating power than DPE (-0.561 eV) and FEME (-0.531 eV), as more negative binding energies correspond to stronger interactions with \ce{Li+}. This trend is also consistent with the higher coordination with \ce{Li+} observed in carbonate solvents compared to ether solvents (Table \ref{tbl:RDF_CN}). Furthermore, among the ethers, FEME exhibits even weaker solvating power than DPE, as reflected by its less negative binding energy.

\begin{gather}
  \label{Eq:DPEFEME_BE}
  E_{\text{b}} = E_{\text{complex}}-(E_{\ce{Li+}} + nE_{\text{solvent}} + mE_{\text{anion}})\\
  \label{Eq:ECDEC_BE}
  E_{\text{b}} = E_{\text{complex}}-(E_{\ce{Li+}} + n_{1}E_{\text{solvent1}} + n_{2}E_{\text{solvent2}} + mE_{\text{anion}})
\end{gather}

\subsubsection{HOMO and LUMO Distributions.~~}

The HOMO and LUMO distributions of all the lithium-ion solvation structures were simulated to investigate their reductive stability and to understand their decomposition mechanisms (Fig. \ref{fig:ESP_CDD_HOMOLUMO}). The LUMO energy level often serves as a key indicator of the reductive stability of electrolyte solutions. Additionally, the molecular orbital diagram of the LUMO can help determine potential decomposition pathways of these solvation structures, as the LUMO is the orbital where electron acceptance occurs during reduction \cite{aoki2022effective}. Furthermore, the energy band gap $(LUMO - HOMO)$ can also determine the chemical reactivity and stability of these solvation structures. A smaller band gap typically corresponds to higher chemical reactivity and lower stability, while a larger band gap suggests reduced reactivity and increased stability \cite{he2022understanding}. Among these solvation structures, the SSIPs without the \ce{PF6-} anion have a lower band gap (0.18 to 0.30 eV) and binding energy (-3.50 to -2.40 eV), making them the most unstable structures (Table \ref{tbl:solvatin_structuer_ECDEC_noPF6}).

The LUMO is primarily distributed on the \ce{FSI-} anion in all \ce{LiFSI}-containing solvation structures in both 1.8 M DPE and 1.8 M FEME electrolytes (Fig. \ref{fig:ESP_CDD_HOMOLUMO}, S20\dag, and S21\dag). Hence, the \ce{FSI-} anions will preferentially undergo reductive decomposition \cite{aoki2022effective}.  In the SSIP, CIP, and AGG structures in EC/DEC+1 M \ce{LiPF6} electrolyte, the LUMO is distributed across both EC and DEC molecules. These findings indicate that reductive decomposition reactions may occur through both EC and DEC decomposition \cite{aoki2022effective}.

\section*{Conclusions}

In the present study, we systematically investigate the electrolyte structures in fluorinated ether (FEME+\ce{LiFSI}), non-fluorinated ether (DPE+\ce{LiFSI}), and organic carbonate-based (EC/DEC+\ce{LiPF6}) electrolytes over a wide range of salt concentrations (1 M, 1.8 M, and 4 M) using a combination of classical MD simulations with the OPLS-AA force field and DFT calculations. We observe that AGGs are the predominant species in the ether-based electrolytes, whereas SSIPs dominate in the mixed carbonate-based electrolyte. This aggregation effect is particularly strong in FEME-based electrolytes, supported by the high coordination number of \ce{Li+-Li+} pairs and the comparatively lower binding energy of FEME to \ce{Li+}. The most dominant solvation structure in each ether-based electrolyte is the anion-rich solvation structure \ce{Li+\ce{(FSI-})3(DPE)1} and \ce{Li+\ce{(FSI-})3(FEME)1}, respectively, and remain nearly unchanged across varying salt concentrations. Regarding the solvent composition in the solvation structures of the EC/DEC electrolyte, a higher fraction of DEC appears to be favorable. Our findings indicate that both DPE and FEME solvents exhibit weak solvating power at all salt concentrations, as indicated by the radial distribution functions, coordination numbers, and solvation structures, which show a strong preference for \ce{Li+} to interact with \ce{FSI-} anions in the primary solvation shell. In particular, FEME shows even weaker solvating power than DPE, as indicated by the higher coordination numbers of \ce{FSI-} in the primary solvation shell of FEME electrolytes. We also observe an increase in unique solvation structures in ether-based electrolytes with higher salt concentrations, with FEME+\ce{LiFSI} displaying a slightly larger variety of structures than DPE+\ce{LiFSI}. Furthermore, the electronic information of the lithium-ion solvation structures obtained from the DFT calculations are quite similar for both DPE- and FEME-based electrolytes. The charge density difference and Bader charge analysis show that the charge transfer from \ce{Li+} to the surrounding solvents and anions in the AGGs of the DPE+1.8 M \ce{LiFSI} electrolyte ($0.70$ - $0.86$ $e$) is comparatively higher than in the FEME+1.8 M \ce{LiFSI} electrolyte ($0.67$ - $0.79$ $e$). The binding energies of these AGGs in the 1.8 M DPE and 1.8 M FEME electrolytes are quite similar, ranging from -8.91 to -7.29 eV and -8.64 to -7.36 eV, respectively. The chemical stability of the solvation structures has also been predicted using their HOMO/LUMO distributions. Fluorinated electrolytes present safety concerns, including volatility and flammability \cite{hou2023thermal}. This study computationally investigates the solvation structure of FEME electrolyte and does not propose for immediate commercial use due to the risk of flammability. Further experimental investigation is needed to gain a deeper understanding of anion-rich solvation structures and flammability in FEME-based electrolytes. Combined with experimental and computational studies, our findings could provide valuable insights for advancing AGG-dominated FEME-based electrolyte design to meet the demands of LiF-rich SEI layers in next-generation lithium-ion batteries.

\section*{Author contributions}

Conceptualization: R.H. Simulation (MD and DFT): R.H. Data Management: R.H. Compilation: R.H. Proofreading: R.H. and D.D. Supervision: D.D. Funding acquisition: D.D.

\section*{Conflicts of interest}

There are no conflicts to declare.

\section*{Data availability}

The input scripts and data files for running MD simulations in LAMMPS, along with the optimized structure files (DPE, FEME, EC, DEC, \ce{LiFSI}, and \ce{LiPF6}) from VASP, are available on the GitHub repository at \href{https://github.com/mana121/SolvationStructure.git}{https://github.com/mana121/SolvationStructure.git}. The supplementary section includes GIF generated using the OVITO software, illustrating the full 5 ns trajectory of the production run for the DPE+1.8 M \ce{LiFSI}, FEME+1.8 M \ce{LiFSI}, and 1:1 EC/DEC+1 M \ce{LiPF6} electrolyte systems. The Supporting Information is available free of charge.

\section*{Acknowledgements}

This work used resources of the the HPC cluster Wulver at the New Jersey Institute of Technology (NJIT) and Expanse, a dedicated Advanced Cyberinfrastructure Coordination Ecosystem: Services and Support (ACCESS) cluster, which are supported by National Science Foundation (NSF) award numbers CMMI-2237990, CBET-2126180, and ACCESS awarded resources or project DMR180013. The authors extend their appreciation to the ViPER Lab of Vilas Pol at Purdue University for their valuable observations on the initial simulation results of electrolytes conducted by the author.



\balance

\renewcommand\refname{References}

\bibliography{rsc} 
\bibliographystyle{rsc} 

\end{document}